%% file: JCTCmain.tex
\documentclass[journal=jctcce,manuscript=preprint]{achemso}
\usepackage{setspace}
\AtBeginDocument{\singlespacing}

\usepackage[version=3]{mhchem} 
\usepackage{lipsum}
\usepackage{accents}
\usepackage{mathrsfs}
\usepackage{physics}
\usepackage{braket}
\usepackage{mathtools}
\usepackage{amsfonts}
\usepackage{amsmath}
\usepackage{caption}
\usepackage{subcaption}
\usepackage{float}
\usepackage{geometry}
\usepackage[numbers]{natbib}

\usepackage[linesnumbered,ruled,vlined]{algorithm2e}
\usepackage{setspace}
\usepackage{xkeyval}
\usepackage{bigints}
\usepackage{multirow}
\usepackage{stackengine}
\usepackage{nicefrac}
\usepackage{titlesec}
\usepackage{booktabs}
\usepackage{varwidth}
\usepackage{hyperref}
\usepackage{enumitem}
\usepackage{esint}
\titleformat{\section}{\large\bfseries}{\thesection}{1em}{}
\usepackage[acronym]{glossaries}
\usepackage[usenames,dvipsnames]{color}
\DeclareMathAlphabet{\mathpzc}{OT1}{pzc}{m}{it}
\usepackage{tcolorbox}
\usepackage{tikz}
\usetikzlibrary{matrix,arrows,positioning,calc}
\DeclareFontFamily{U}{mathx}{\hyphenchar\font45}
\DeclareFontShape{U}{mathx}{m}{n}{
      <5> <6> <7> <8> <9> <10>
      <10.95> <12> <14.4> <17.28> <20.74> <24.88>
      mathx10
      }{}
\DeclareSymbolFont{mathx}{U}{mathx}{m}{n}
\DeclareFontSubstitution{U}{mathx}{m}{n}
\DeclareMathAccent{\widecheck}{0}{mathx}{"71}
\DeclareMathAccent{\wideparen}{0}{mathx}{"75}
\DeclareMathAccent{\widebar}{0}{mathx}{"73}

\titleformat{\subsection}
  {\normalfont\bfseries\normalsize}
  {\thesubsection}
  {0.5em}
  {}

\titleformat{\subsubsection}
  {\normalfont\itshape\footnotesize}
  {\thesubsubsection}
  {0.5em}
  {}

\newcommand{\pawfe}{\texttt{PAW-FE}}
\newcommand{\bR}{\boldsymbol{\textbf{R}}}
\newcommand{\br}{\boldsymbol{\textbf{r}}}

\newcommand{\bT}{\boldsymbol{\textbf{T}}}
\newcommand{\bJ}{\boldsymbol{\textbf{J}}}
\newcommand{\bx}{\boldsymbol{\textbf{x}}}
\newcommand{\bX}{\boldsymbol{\textbf{X}}}
\newcommand{\bY}{\boldsymbol{\textbf{Y}}}

\newcommand{\by}{\boldsymbol{\textbf{y}}}
\newcommand{\bk}{\boldsymbol{\textbf{k}}}
\newcommand{\bL}{\boldsymbol{\textbf{L}}}
\newcommand{\bH}{\boldsymbol{\textbf{H}}}
\newcommand{\bI}{\boldsymbol{\textbf{I}}}
\newcommand{\bS}{\boldsymbol{\textbf{S}}}

\newcommand{\bF}{\boldsymbol{\textbf{F}}}

\newcommand{\bA}{\boldsymbol{\textbf{A}}}

\newcommand{\bQ}{\boldsymbol{\textbf{Q}}}
\newcommand{\bZ}{\boldsymbol{\textbf{Z}}}
\newcommand{\bP}{\boldsymbol{\textbf{P}}}
\newcommand{\bM}{\boldsymbol{\textbf{M}}}
\newcommand{\bD}{\boldsymbol{\textbf{D}}}

\newcommand{\bC}{\boldsymbol{\textbf{C}}}
\newcommand{\bV}{\boldsymbol{\textbf{V}}}

\newcommand{\qe}{\texttt{QE}}
\newcommand{\abinit}{\texttt{Abinit}}
\newcommand{\feNodalVectorUtilde}{\widetilde{\boldsymbol{\mathsf{u}}}}

\newcommand{\Psips}{\widetilde{\boldsymbol{\Psi}}}
\newcommand{\phiae}{\phi}

\newcommand{\phips}{\widetilde{\phi}}
\newcommand{\nTilde}{\widetilde{n}}
\newcommand{\mTilde}{\widetilde{m}}
\newcommand{\rhoTilde}{\widetilde{\rho}}
\newcommand{\uTilde}{\widetilde{u}}
\newcommand{\gTilde}{\widetilde{g}}
\newcommand{\pTilde}{\widetilde{p}}
\newcommand{\bTilde}{\widetilde{b}}

\newcommand{\fePsips}{\widetilde{u}}
\newcommand{\bfePsips}{\widetilde{\boldsymbol{u}}}

\newcommand{\DFTFE}{\texttt{DFT-FE}}

\newcommand{\cn}{\color{black}}
\newcommand{\cb}{\color{black}}

\newcommand{\imag}{\mathrm{i}}

\author{Kartick Ramakrishnan}
\affiliation[CDS Indian Institute of Science, Bangalore]
{Department of Computational and Data Sciences, Indian Institute of Science, Bengaluru 560012, India}
\author{Phani Motamarri}
\email{phanim@iisc.ac.in}
\affiliation[CDS Indian Institute of Science, Bangalore]
{Department of Computational and Data Sciences, Indian Institute of Science, Bengaluru 560012, India}

\title
{\Large Accelerating finite-element-based projector augmented-wave DFT Calculations with scalable GPU-centric computational methods}
\abbreviations{}

\begin{document}


\begin{tocentry}
\includegraphics[width=\linewidth]{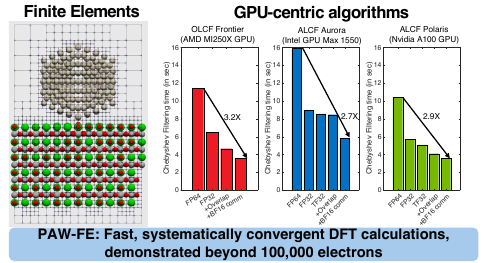} 
\end{tocentry}



\begin{abstract}

\cb  First principles calculations at scale of $10^4$--$10^5$
electrons for complex material systems, with practical time-to-solution, remain a significant challenge for existing density functional theory (DFT) implementations. In this work, we fill this gap by presenting GPU-centric computational methods and algorithmic innovations for the finite-element (FE) discretization of projector augmented-wave (PAW) formulation of DFT (\pawfe) that enable accurate, efficient, and scalable electronic-structure calculations with generic boundary conditions on modern exa-scale architectures. The resulting PAW generalized Hermitian eigenproblem is solved using the Residual-based Chebyshev Filtered Subspace Iteration (R-ChFSI) method tolerant to inexact matrix-multivector products. Exploiting this tolerance to matrix-vector products, we employ an approximate inverse PAW overlap matrix, mixed-precision arithmetic (\texttt{FP32}/\texttt{TF32}), low-precision nearest-neighbor communication (\texttt{BF16}), and block-wise computation--communication overlap to reduce computational cost while preserving numerical accuracy. These algorithmic ideas deliver overall CPU-GPU speedups of $8\times$ on ALCF Aurora (Intel GPUs), $20\times$ on OLCF Frontier (AMD GPUs) and $14\times$ on ALCF Polaris (Nvidia GPUs). Relative to widely used plane-wave PAW implementations, \pawfe~achieves approximately an $7\times$ reduction in time-to-solution for systems containing around 10,000 electrons on NVIDIA GPUs, with much higher gains at larger system sizes, and a $6\times$ speedup over norm-conserving pseudopotential finite-element DFT implementations. We further demonstrate strong scalability for material systems comprising $130,000$ electrons, establishing \pawfe~as an exascale-ready framework for large-scale first-principles simulations with systematically controllable numerical accuracy.
\end{abstract}
\printglossary[type=\acronymtype]

\section{Introduction}
\input{intro}

\section{Mathematical formulation}\label{sec: 2}
\input{mathformulation.tex}

\section{Computational Methodology}\label{sec: 3}
\input{computationalMethodology}
\section{Numerical Implementation}\label{sec: 4}
\input{NumericalImplementation}
\section{Results}\label{sec: 5}
\input{results.tex}

\section{Perspectives and concluding remarks}\label{sec: 6}
\input{Discussion.tex}



\begin{acknowledgement}


\input{acknowledgement}

\end{acknowledgement}
\newpage
\section*{Data and Software availability}
All input, output  files and source code that yielded the results presented in this work are available freely online in our \href{https://github.com/matrixlabiisc/PAW-FE_GPU}{GitHub} repository (https://github.com/matrixlabiisc/PAW-FE\_GPU).

\begin{suppinfo}
Details of the PAW governing equations, the self-consistent field iteration algorithm, efficient evaluation of the FE-discretized operator action, subspace construction algorithms, and tables of accuracy benchmarking results are provided in the Supporting Information.
\end{suppinfo}
\bibliography{JCTC}
 \pagebreak


\end{document}

%% file: intro.tex
Electronic-structure calculations have played a pivotal role in accurately predicting diverse material properties and today stand as the primary workhorse in modern materials discovery. Applications of electronic-structure calculations based on Kohn–Sham density functional theory (DFT)~\cite{KSDFT, HKNoble, martin2020electronic} have found remarkable success in identifying viable materials for energy storage systems, thermoelectric devices, predicting dopants and alloy compositions that enhance electrical, mechanical, and optical properties, and providing quantitative insight into the mechanisms that govern charge transport, defect chemistry, and catalytic reaction pathways. This ubiquitous adoption is largely due to the computational tractability afforded by DFT, which enables simulations of complex material systems while retaining a close connection to the underlying quantum-mechanical description of electronic interactions. Widely used plane-wave implementations such as VASP,\cite{PhysRevB.54.11169} Quantum ESPRESSO,\cite{qe} and ABINIT\cite{Abinit2016} have further enabled high-throughput materials screening and the generation of reference datasets for training machine-learning interatomic potentials, including MACE~\cite{MACE}, Allegro~\cite{Allegro} and NEP~\cite{NEP}.

Despite this success, many scientifically important problems now demand simulations at length scales that significantly exceed those accessible with conventional approaches. Computing spin textures such as magnetic skyrmions, modeling catalytic reactions on substrate-supported metal nanoparticles, studying defect formation and migration in crystalline solids, and describing electrode–electrolyte interfaces in solid-state batteries all require large simulation cells to eliminate finite-size artifacts, capture charge transfer and strain effects, avoid spurious defect–defect interactions, and represent chemically complex environments spanning several nanometers and involving thousands of atoms. In all these settings, an accurate electronic-structure description is essential, calling for simulations of $10,000$-$100,000$ electrons with \cb numerical discretization errors below targeted chemical accuracy threshold\cn. Such large-scale calculations are equally indispensable for validating machine-learned interatomic potentials (MLIPs) at length scales beyond those sampled during training.

\cb Although sub-cubic and linear-scaling formulations of Kohn–Sham DFT exist~\cite{PEXSI}, their additional approximations and large crossover system sizes leave diagonalization-based cubic-scaling solvers as the method of choice for the complex heterogeneous material systems targeted here, where \cn the practical barriers to routine large-scale calculations are of a different nature: high time-to-solution, large memory footprints arising from the need for many basis functions per atom to achieve desired numerical accuracy, poor parallel scalability on modern multi-node architectures, and eigensolver algorithms ill-suited to exploit the latest generation GPU hardware. The Projector Augmented-Wave (PAW) method introduced by Blöchl~\cite{PAW} addresses a few of these limitations by reformulating the all-electron problem, characterized by rapidly oscillatory valence wavefunctions near atomic cores, into an equivalent problem involving smooth pseudo-wavefunctions augmented by localized atom-centered corrections. Within the frozen-core approximation, this transformation allows valence states to be represented with significantly fewer basis functions while preserving all-electron accuracy, thereby substantially reducing both the dimensionality of the discretized Kohn–Sham problem and its associated memory footprint. Plane-wave implementations such as VASP~\cite{PhysRevB.54.11169}, Quantum ESPRESSO~\cite{qe}, and ABINIT~\cite{Abinit2016} have widely adopted the PAW method~\cite{PAW, Kresse1999}, making them the predominant tools in the materials science community. However, plane-wave implementations inherently enforce periodic boundary conditions and rely on global fast Fourier transforms (FFTs), which incur substantial all-to-all communication. On modern GPU-accelerated supercomputers, where floating-point throughput has grown dramatically relative to memory bandwidth and inter-node communication capacity, FFT-based implementations are increasingly bandwidth-bound and communication-limited, posing a fundamental obstacle to scalable large-scale DFT on exascale systems.

Real-space discretizations provide an alternative route that is better suited to leverage multi-node computing architectures. In addition to improved parallel scalability, real-space methods allow the use of generic boundary conditions, which provide a more realistic computational setup for modeling slabs, surfaces, and nanoclusters~\cite{electricFieldJCTC}. Real-space methods include finite difference approaches such as Octopus\cite{octopus2015}, GPAW\cite{GPAW2010}, PARSEC\cite{PARSEC}, SPARC\cite{sparc2017a}, wavelet based approaches in BigDFT\cite{bigdft}  and finite-element approaches in DFT-FE~\cite{dftfe0.6,dftfe1.0,Das2019FastSystem, Das2023}, FEMTECK~\cite{tsuchida1995,Tuschida}. 
However, with the exception of GPAW\cite{GPAW2010}, most existing real-space implementations rely on the norm-conserving pseudopotential (ONCV) approximation to solve the DFT problem.  Although systematically transferable, ONCV pseudopotentials are hard, particularly for first-row and transition-metal elements, and therefore require a substantially larger number of basis functions to achieve \cb required discretization error in energies and forces \cn. This leads to a substantial increase in computational cost and memory footprint. In contrast, the projector augmented-wave (PAW) method reconstructs the true all-electron wavefunctions from smooth pseudo-wavefunctions through an explicit linear transformation. This formulation enables accurate recovery of the all-electron charge density and wavefunctions in the vicinity of atomic nuclei, allowing reliable evaluation of quantities sensitive to core regions such as forces, hyperfine interactions, and electrostatic potentials, while retaining the efficiency of a smooth basis representation. Furthermore, by relaxing the strict norm-conservation constraint, PAW typically achieves faster basis set convergence and improved transferability compared to norm-conserving pseudopotentials.

Real-space finite-element (FE) discretization for the projector augmented-wave method (\texttt{PAW-FE}\cite{pawfe}) has only recently been explored, where high-order spectral finite elements are employed to solve the PAW governing equations. This approach demonstrates a natural synergy between the smooth fields arising from the PAW formalism and the adaptive resolution and systematic convergence of finite-element discretizations. In \texttt{PAW-FE}, higher-order finite-element basis functions with compact support are used to discretize the governing equations, enabling systematic polynomial convergence while naturally accommodating periodic, semi-periodic, and non-periodic boundary conditions. The locality of these FE basis functions induces sparse operator structure requiring only nearest-neighbor communication~\cite{pawfe, dftfe1.0} to evaluate its action, thereby enabling fine-grained parallelism well suited for large-scale simulations on exascale architectures. The FE-discretized PAW governing equations derived in our previous work~\cite{pawfe} lead to a generalized Hermitian eigenvalue problem (GHEP) of the form $\bH \Psips = \bS \Psips \boldsymbol{\Lambda}$. To efficiently solve this problem, this previous work extended the traditional Chebyshev filtered subspace iteration (ChFSI) eigensolver, originally employed for standard Hermitian eigenvalue problems, to GHEP. The resulting PAW-FE methodology implemented on distributed multinode CPU architectures demonstrated excellent agreement with plane-wave PAW implementations across representative benchmark systems with non-periodic, semi-periodic, and fully periodic boundary conditions, including comparisons of ground-state energies, energy variation against bond-length and lattice parameters, formation energies, surface energies, and band structures. In addition to accuracy, the proposed computational strategies enabled substantial performance gains relative to plane-wave implementations on multinode CPU architectures. Specifically, PAW-FE achieved approximately $5\times$ reduction in computational cost for non-periodic systems containing $\sim18{,}000$ electrons and about $3\times$ reduction for periodic systems containing $\sim34{,}000$ electrons, with greater gains observed as system size increased. Furthermore, significant reductions in minimum wall time ($\sim7\times$–$10\times$) were observed for both periodic and non-periodic systems. The method also exhibited excellent parallel scaling efficiency on multinode CPU architectures ($\sim70\%$) even at 1,200 degrees of freedom per MPI task and achieved nearly an order-of-magnitude reduction in computational cost compared to the norm-conserving pseudopotential implementation in the \DFTFE~code for systems approaching 50,000 electrons.

Motivated by the promising results of \pawfe\ on multi-node CPU architectures, this work develops new eigensolver algorithms and numerical implementation strategies for the PAW generalized Hermitian eigenproblem (GHEP) that are well-suited to modern multi-node GPU architectures. Contemporary GPU hardware, which is largely optimized for machine-learning workloads, delivers extremely high throughput through low-precision tensor cores (\texttt{FP16}/\texttt{BF16}/\texttt{TF32}), yet continues to face bottlenecks associated with memory bandwidth and data movement. Achieving optimal performance on such hardware, therefore, requires minimizing data movement, maximizing arithmetic intensity, overlapping communication with computation, and leveraging reduced-precision arithmetic without sacrificing numerical robustness. Finite-element (FE) discretizations are particularly well-positioned to meet these demands, as the cell-structured representation of both the Hamiltonian and the PAW overlap operator enables their action to be reformulated as dense cell-level matrix–matrix operations, naturally reducing global synchronization and facilitating computation–communication overlap. Central to our approach is an extension of the recently proposed Residual-based Chebyshev Filtered Subspace Iteration (R-ChFSI) algorithm~\cite{RChFSI} to solve the PAW-GHEP efficiently on multi-node GPU architectures. The tolerance of R-ChFSI to inexact matrix–vector products during Chebyshev-filtered subspace construction enables two key algorithmic innovations. First, it permits the use of an approximate inverse PAW overlap matrix, eliminating the need for an inner iterative solver as proposed in PW codes~\cite{LEVITT201598}. Second, it enables the use of reduced-precision arithmetic (\texttt{FP32}/\texttt{TF32}) and low-precision nearest-neighbor communication (\texttt{BF16}) during the Chebyshev filtering step. 
\cb Leveraging these algorithmic capabilities, the present work extends our previous CPU-based PAW-FE ~\cite{pawfe} through the following methodological, GPU-centric algorithms and implementation advances:
(i) formulation of the FE-discretized governing equations and the corresponding ionic forces for the PAW method within a collinear spin formalism;
(ii) development of a multi-resolution quadrature scheme for the accurate evaluation of atom-centered integrals arising in the FE discretization of PAW matrices on coarse grids;
(iii) use of an R-ChFSI-based eigensolver for the nonlinear FE-discretized PAW generalized Hermitian eigenproblem, eliminating the need for explicit inversion of the PAW overlap operator;
(iv) design of a computation--communication overlap strategy within the Chebyshev-filtered blocked subspace construction; and
(v) introduction of mixed-precision computation (\texttt{FP32}/\texttt{TF32}) together with low-precision MPI communication (\texttt{BF16}) during the Chebyshev filtering step, without any significant loss in accuracy while achieving nearly a three-fold reduction in the cost of Chebyshev-filtered subspace construction.
Collectively, these advances enable \pawfe~to efficiently exploit modern exascale GPU architectures for scalable large-scale first-principles calculations.\cn

We demonstrate that the proposed approach achieves  \cb the desired numerical accuracies in energies and forces \cn across periodic and non-periodic benchmark systems while substantially reducing time-to-solution relative to plane-wave PAW methods at comparable accuracy. Ground-state calculations involving approximately $10^{4}$ electrons can be completed within minutes on modern multi-GPU nodes, with strong parallel scaling maintained across large GPU nodes. These results indicate that real-space \pawfe~ methods can routinely extend DFT simulations into the $10^{4}$-- $10^{5}$ electron regime with practical wall times. By combining the reduced degrees of freedom enabled by PAW, the systematic convergence and locality of high-order finite elements, and GPU-centric eigensolver design, this work establishes an exascale-ready method for large-scale, chemically accurate first-principles simulations. Specifically, \pawfe~achieves approximately $3.6\times$ reduction in computational cost compared to Quantum ESPRESSO, a plane-wave based code, for system sizes around 10,000 electrons on NVIDIA A100 GPUs with much larger gains achievable for system sizes greater than 10,000 electrons. Further, we observe a reduction of around $7\times$ in the minimum time-to-solution for system sizes approaching 10,000 electrons and of around $3\times$ for system sizes around 3,000 electrons. Furthermore, when compared against \DFTFE, it is observed that \pawfe~requires $4\times$ fewer compute nodes and achieves $5\times$ reduction in computational cost. Finally, we consider twisted bilayer WTe\textsubscript{2} comprising 130,000 electrons (11,000 atoms) as a representative large-scale system to demonstrate the capability of the proposed approach for realistic materials applications, which is beyond the reach of current DFT codes that employ systematically convergent basis sets.

The rest of the manuscript is organized as follows. Section~\ref{sec: 2} presents in brief the governing equations of a local real-space formulation of the PAW method within the collinear spin formalism. The section further introduces the finite-element discretization of the governing equations.
Section~\ref{sec: 3} presents the proposed computational methodology for achieving an efficient and scalable solution of the FE-discretized PAW generalized eigenproblem using a self-consistent field iteration framework. Section~\ref{sec: 4} details the numerical implementation strategies for efficient computation of the ground state on modern GPU architectures.
Section~\ref{sec: 5} provides comprehensive accuracy and performance benchmarks of the proposed implementation by comparing against state-of-the-art plane-wave PAW codes on representative benchmark systems. Additionally, this section presents a comparative study of the present PAW-FE approach and norm-conserving pseudopotential (ONCV) calculations performed using the DFT-FE code, focusing on parallel scalability and time-to-solution. Finally, Sec.~\ref{sec: 6} summarizes the key findings of this work and outlines future research directions arising from the proposed methodology.

%% file: mathformulation.tex
 
We begin by introducing the key equations of spin-polarised density functional theory (DFT) within the projector augmented-wave (PAW) formalism in a real-space setting. The detailed formulation is presented in the Supplementary Information (sec. S-1). Following this,  we present the finite-element discretization of the PAW governing equation.

\subsection{DFT governing equations in the PAW method}
For a material system comprising `$N_a$' atoms and `$N_v$' valence electrons in the periodic unit-cell ($\Omega_p$) with atomic positions denoted by $\{ \bR^a \}$ ($\bR^{a}$ refers to the atomic position vector of the $a$\textsuperscript{th} atom), the ground-state solution in Bl\"{o}chl's projector augmented-wave (PAW) formalism~\cite{PAW} within the frozen-core approximation is determined by solving the $N \geq \frac{N_v}{2}$ smallest eigenstates of the following generalised Hermitian eigenproblem (PAW-GHEP) : 
\begin{equation}  \boldsymbol{\mathcal{H}}^{\sigma,\bk}\uTilde^\sigma_{n\bk} = \varepsilon^{\sigma}_{n\bk}\boldsymbol{\mathcal{S}}^{\bk}\uTilde^\sigma_{n\bk},\;\; \forall \sigma \in \{ \uparrow, \downarrow \},\; 1 \leq n \leq N
\label{eq: PAW-GHEP}
\end{equation}
where $\sigma \in \{ \uparrow, \downarrow \}$ denotes the spin-index, and  $\uTilde^\sigma_{n\bk}(\bx)$ is a complex-valued pseudo (PS) wavefunction that is periodic on the unit cell satisfying $\uTilde^\sigma_{n\bk}(\bx+\bL_{r})=\uTilde^\sigma_{n\bk}(\bx)$ for all lattice vectors $\bL_{r}$, and $\bk$ denotes a point in the first Brillouin zone of the underlying reciprocal lattice. The PAW Hamiltonian operator $\boldsymbol{\mathcal{H}}^{\sigma,\bk} = \boldsymbol{\mathcal{H}}^{\sigma,\bk}_\text{loc} + \boldsymbol{\mathcal{H}}^{\sigma,\bk}_\text{nloc} $ has both local and non-local contribution. The local contribution of the Hamiltonian is given by
\begin{equation}
  \boldsymbol{\mathcal{H}}^{\sigma,\bk}_\text{loc} = -\frac{1}{2}\nabla^2 -\imag \bk \cdot \nabla + \frac{1}{2}|\bk|^2+ V^\sigma_{\text{eff}}(\rhoTilde,\mTilde) + V_{\text{z}}
  \label{eq: PAW Hamiltonian}
\end{equation}
where the effective potential $V^\sigma_{\text{eff}}(\rhoTilde,\mTilde)$ is defined as
\begin{equation}
    V^\sigma_{\text{eff}}(\rhoTilde,\mTilde) :=  V_{\text{xc}}(\nTilde^\sigma(\bx)) + V_{\text{el}}(\rhoTilde(\bx)) 
    \label{eq: Veff}
\end{equation}
and depends on the pseudo total charge density $\rhoTilde(\bx)$ and 
pseudo magnetization density $\mTilde(\bx)$, through the spin density $\nTilde^\sigma(\bx)$. We note that $\rhoTilde(\bx) = \nTilde(\bx)+\bTilde(\bx)$ where $\nTilde(\bx) = \sum_\sigma \nTilde^{\sigma}(\bx)$ denotes the pseudo electron density and $\bTilde(\bx)$ is the total compensation charge density, where both of which depend on the pseudo wavefunction $\{\uTilde^{\sigma}_{n\bk}\}$. We refer the reader to the Supplementary Information for a detailed discussion of the expressions for the various terms involved in eqn.~\eqref{eq: Veff} (see S-1). Further, the action of the non-local Hamiltonian is given by
\begin{equation} \boldsymbol{\mathcal{H}}^{\bk,\sigma}_{\text{nloc}}\uTilde^{\sigma}_{n\bk} := \sum_{a}\sum_{\alpha\beta}\sum_{rr'}\pTilde^a_{\alpha}(\bx-\bR^a-\bL_{r})e^{-\imag \bk\cdot(\bx-\bL_r)}\Delta h^{a,\sigma}_{\alpha\beta}\int_{\Omega_p}{ \pTilde^a_{\beta}(\by-\bR^a-\bL_{r'})e^{\imag\bk\cdot(\by-\bL_{r'})}\uTilde^\sigma_{n\bk}(\by) d\by} 
\end{equation}
where $\Omega_p$ refers to the periodic unit-cell. Similar to the Hamiltonian operator, the PAW overlap operator $\boldsymbol{\mathcal{S}}^{\bk}$ has both local and nonlocal contributions, with the local contribution being the identity operator $\boldsymbol{\mathcal{I}}$,  and the action of the non-local operator is given by
\begin{equation} \boldsymbol{\mathcal{S}}^{\bk,}_{\text{nloc}}\uTilde^{\sigma}_{n\bk} := \sum_{a}\sum_{\alpha\beta}\sum_{rr'}\pTilde^a_{\alpha}(\bx-\bR^a-\bL_{r})e^{-\imag \bk\cdot(\bx-\bL_r)}\Delta s^{a}_{\alpha\beta}\int_{\Omega_p}{ \pTilde^a_{\beta}(\by-\bR^a-\bL_{r'})e^{\imag\bk\cdot(\by-\bL_{r'})}\uTilde^\sigma_{n\bk}(\by) d\by} 
\end{equation}
with the expressions to evaluate the coupling matrices ($\Delta h^{a,\sigma}_{\alpha\beta}$, $\Delta s^{a}_{\alpha\beta}$) are presented in the Supplementary Information (see S-1).

The governing equations for determining the DFT ground-state  requires the solution of a non-linear generalised eigenvalue problem that needs to be solved self-consistently in conjunction with a Poisson problem for the electrostatics (see Eqn. S-18 of Supplementary Information). To this end, solving the ground-state in PAW formalism can be posed as the following fixed-point problem:
\begin{equation}
  \{\rhoTilde_{\text{out}}(\bx), \mTilde_{\text{out}}(\bx),\{\bD_{\text{out}}^{a,\text{mag}} \}  \} = F[\{\rhoTilde_{\text{in}}(\bx), \mTilde_{\text{in}}(\bx),\{\bD_{\text{in}}^{a,\text{mag}} \}  \}] 
\end{equation}
where the evaluation of $F[\cdot]$ involves: (i)
computing the effective potential given in Eqn.~\eqref{eq: Veff}, the coupling matrices $\Delta h^{a,\sigma}_{\alpha\beta}$ in Eqn. S-13 (refer to supplementary information) both of which in turn depend on the input fields $\{\rhoTilde_{\text{in}}(\bx), \mTilde_{\text{in}}(\bx),\{\bD_{\text{in}}^{a,\text{mag}} \}  \}$, (ii) solving the PAW-GHEP in Eqn.~\eqref{eq: PAW-GHEP}, and finally (iii)  computing $\{\rhoTilde_{\text{out}}(\bx), \mTilde_{\text{out}}(\bx),\{\bD_{\text{out}}^{a,\text{mag}} \}  \}$ from the eigenvalues and eigenfunctions as illustrated in Eqns. S-4, S-8, S-10 of Supplementary Information section. As detailed in our previous work~\cite{pawfe}, we employ Anderson mixing~\cite{Anderson1965} to accelerate the convergence of the aforementioned Kohn-Sham fixed point problem by extending this mixing scheme to include the magnetization density. Specifically, the Anderson mixing coefficients are determined by minimising the objective 
\begin{equation}
    R^2 = ||\underline{\rhoTilde}_{\text{in}}- \underline{\rhoTilde}_{\text{out}}||^2_W+ K||\underline{\mTilde}_{\text{in}}-\underline{\mTilde}_{\text{out}}||_2^2 + A \sum_a || \underline{\bD}^{a,\text{mag}}_{\text{in}} - \underline{\bD}^{a,\text{mag}}_{\text{out}}||^2_{\text{F}}
    \label{eq: anderson minimizer}
\end{equation}
 with $K$ denoting the spin-enhancement factor, which is chosen to be around $4.0$\cite{VASP_AMIX_MAG_Wiki} and $A$ is typically chosen to be around $1\times10^{-4}$ for most systems. In Eqn.~\eqref{eq: anderson minimizer}, $||.||_2$ refers to the L\textsubscript{2}-norm of the field, $||.||_{\text{F}}$ refers to the matrix Frobenius norm, and $||.||_{W}$ refers to the weighted norm that is employed to reduce the sloshing behaviour of electron density, especially in metallic systems, by dampening out the long wavelength oscillations of electron density residuals across SCF iterations.  The real-space evaluation details of $||.||_{W}$ is discussed in our prior work~\cite{pawfe}.
 Furthermore, $\underline{\rhoTilde}_{\text{in(out)}}$ and  $\underline{\mTilde}_{\text{in(out)}}$ represent the input and output total pseudo charge density and pseudo magnetisation density, respectively, while $\underline{\bD}^{a,\text{mag}}_{\text{in(out)}} $ represents the input and output atom-dependent magnetization density matrix, all of which are expressed as linear combinations of the input and output fields from previous SCF iterations. These linear combination coefficients are obtained by minimising the objective defined in Eqn.~\eqref{eq: anderson minimizer}.  We note that, in contrast to the weighted $L_2$ norm employed for total charge density, only $L_2$ norm is used for the magnetization density residual $\mTilde_{\text{res}}$, as it does not contribute to sloshing. This follows from the fact that the 
magnetization density $\mTilde(\bx)$ enters the Kohn–Sham equations exclusively through the exchange–correlation potential
$V^\sigma_{\text{xc}}$, which is semi-local in nature within the generalized gradient approximation.
 
Finally, at SCF convergence, the all-electron energy is computed using the double-counting expression
\begin{align}
    E_0 =& \sum_{\sigma}\sum_n{\fint_{\text{BZ}}{f^\sigma_{n\bk}}\varepsilon^\sigma_{n\bk}}d\bk - \int_{\Omega_p}{\sum_{\sigma}\left(V_{\text{el}}^{\text{in}}(\bx)+V^{\sigma,\text{in}}_{\text{xc}}+V_{\text{z}}\right)\nTilde^\sigma_{\text{out}}(\bx) d\bx} - \sum_{\sigma}\sum_a{\Delta H^{\sigma a,\text{in}}_{\alpha\beta}D^{\sigma a,\text{out}}_{\alpha\beta}} \nonumber \\
    &+ E_\text{xc}\left[\nTilde_{\text{out}}(\bx),\mTilde_{\text{out}}(\bx)\right] + E_{\text{el}}\left[\rhoTilde_{\text{out}}(\bx)\right ] +  \sum_{\sigma}\sum_a{\Delta E^{\sigma,a}[\left\{D^{\sigma,a,\text{out}}_{\alpha\beta}\right\}]} 
    \label{eq: double counting}
\end{align}
Finally, the ionic forces, which are the derivatives of the ground-state energy with respect to atom positions as $\bF^a = -\frac{\partial E}{\partial \bR^a}$ have the following expressions:
\begin{tcolorbox}[colback=white,colframe=black,boxrule=0.8pt,
  arc=2mm, width=\dimexpr\textwidth+2cm\relax, enlarge left by=-1cm]
  \begin{equation}
    \bF^a = \bF^a_{\text{xc}} + \bF^a_{\text{core}} +  \bF^a_{\text{comp}} + \bF^a_{\text{zero}} + \bF^a_{\text{nl}} + \bF^{a^{*}}_{\text{nl}}
    \label{eq: ionic forces PAW}
\end{equation}
\begin{align}
\bF^a_{\text{xc}} &= -\sum_{r}{\int_{\Omega_p}{\frac{\partial \epsilon_{\text{xc}}}{\partial \nTilde(\bx)}\nabla \nTilde^a_c(\bx-\bR^a-\bL_{r}) d\bx}} - \sum_{r}\int_{\Omega_p}{\left[\nabla\left(\nabla \nTilde^a_c(\bx-\bR^a-\bL_{r})\right)\right]\frac{\partial \epsilon_{\text{xc}}}{\partial \nabla \nTilde} d\bx} \nonumber\\
\bF^a_{\text{core}} &= \sum_{r}{\int_{\Omega_p}{\left[\nTilde^a_c(\bx-\bR^a-\bL_{r})\right]\nabla V_{\text{el}}(\bx)d\bx}},\;\;\; 
\bF^a_{\text{comp}} ={\sum_{r}{\int_{\Omega_p}{\left[\bTilde^a(\bx-\bR^a-\bL_{r})\right]\nabla V_{\text{el}}(\bx)d\bx}}}  \nonumber\\
\bF^a_{\text{zero}} &= \sum_{r}\int_{\Omega_p}{\sum_{\sigma}\sum_n(\nabla |\uTilde^{\sigma}_{n\bk}(\bx)|^2)V^a_{\text{z}}(\bx-\bR^a-\bL_r)d\bx} \nonumber \\
\bF^a_{\text{nl}} &= \sum_{\sigma}\sum_n\sum_{\alpha\beta}\fint_{BZ}f^{\sigma}_{n\bk}\sum_{rr'}\int_{\Omega_p}\pTilde^a_{\alpha}(\bx-\bL_{r'}-\bR^a)e^{-i\bk\cdot(\bx-\bL_{r'})}\nabla\uTilde^{\sigma*}_{n\bk}(\bx)d\bx\,\left( \Delta h^{a,\sigma}_{\alpha\beta} - \varepsilon^{\sigma}_{n\bk}\Delta s^a_{\alpha\beta} \right)\nonumber\\
&\quad\quad\int_{\Omega_p}\pTilde^a_{\beta}(\by+\bL_{r}-\bR^a)e^{\imag\bk\cdot(\by+\bL_{r})}\uTilde^{\sigma}_{n\bk}(\by)d\by \nonumber \\
&+\sum_{\sigma}\sum_{\alpha\beta}\sum_n\fint_{BZ}f_{n\bk}\sum_{rr'}\int_{\Omega_p}\pTilde^a_{\alpha}(\bx-\bL_{r'}-\bR^a)e^{-\imag\bk\cdot(\bx-\bL_{r'})}\left\{-\imag\bk\uTilde^{\sigma*}_{n\bk}(\bx)\right\}d\bx\,\left( \Delta h^{a,\sigma}_{\alpha\beta} - \varepsilon^{\sigma}_{n\bk}\Delta s^a_{\alpha\beta} \right)\nonumber \\
&\quad\quad\int_{\Omega_p}\pTilde^a_{\beta}(\by+\bL_{r}-\bR^a)e^{\imag\bk\cdot(\by+\bL_{r})}\uTilde^{\sigma}_{n\bk}(\by)d\by \nonumber
\end{align}
\end{tcolorbox}

Furthermore, the all-electron density ($n(\bx)$) and the all-electron Kohn-Sham wavefunctions ($\{u^{\sigma}_{n\bk}\}$), which are required for post-processing the ground-state solution for information like projected DOS, electron localization functions (ELF), projected Hamilton populations (p-COHP), can be recovered from the smooth PAW ground-state solution fields as
\begin{align}
    n^{\sigma}(\bx) &= \nTilde^{\sigma}(\bx) + \sum_{r}\sum_a \left( n^{a,\sigma}(\bx-\bR^a-\bL_{r}) - \nTilde^{a,\sigma}(\bx - \bR^a-\bL_{r}) \right)  \label{eq: AE-density}\\
    u_{n\bk}^{\sigma}(\bx) &= \uTilde^{\sigma}_{n\bk}(\bx) + \sum_r\sum_a\sum_{\alpha}\left(\phiae_{\alpha}^a(\bx-\bR^a-\bL_r)-\phips^a_{\alpha}(\bx-\bR^a-\bL_r) \right)\braket{\pTilde^a_{\alpha}|\uTilde^{\sigma}_{n\bk}}
    \label{eq: AE-wfc}
\end{align}
where $n^{a,\sigma}(\bx-\bR^a)$ and $\nTilde^{a,\sigma}(\bx-\bR^a)$ denote the atom-centered all-electron and pseudo electron densities, respectively,  while $\{\phi_{\alpha}^a\}$ and $\{\phips^a_{\alpha}\}$ denote the corresponding atom-dependent all-electron and pseudo partial waves. Further details of these quantities are provided in the Supplementary Information.
\subsection{Finite-element discretization}\label{sec: FE disc}
 The finite-element method involves decomposing the spatial domain into non-overlapping subdomains called finite elements (or cells). The underlying finite-element basis is constructed over Gauss-Legendre-Lobatto nodal points as is usually done in spectral finite element methods~\cite{motamarri2013, pawfe}. FE basis functions chosen in this work are $\mathcal{C}^0$-continuous Lagrange polynomials that are compactly supported in cells sharing a nodal point. Discretizing the electronic fields involved in Eqns.~\eqref{eq: PAW-GHEP}-\eqref{eq: Veff} , we have
\begin{align} \label{eqn:FEbasisexp}
    \fePsips^{\sigma,h}_{i,\bk}(\bx) = \sum_{J=1}^{M}{N^{h,p}_J(\bx)\fePsips^{\sigma,J}_{i,\bk}},\; V_{\text{el}}^{h,p_{\text{el}}}(\bx) = \sum_{J=1}^{M_{\text{el}}}{N^{h,p_{\text{el}}}_J(\bx)V_{\text{el}}^J}
\end{align}
where $\fePsips^{\sigma,J}_{i,\bk}$, $V_{\text{el}}^J$ denote the nodal vectors representing the FE discretised wavefunctions and the electrostatic potential respectively, while $N^{h,p}_J(\bx):1 \leq J \leq M$ and $N^{h,p_{\text{el}}}_J(\bx):1 \leq J \leq M_{\text{el}}$ denote the strictly local Lagrange polynomial basis functions of degree $p$ and $p_{\text{el}}$, respectively spanning the finite-element subspace, generated using the FE triangulation $\mathcal{T}^h$ with the characteristic mesh size denoted by $h$, with $p_{\text{el}} \geq p$. To this end, the finite-element discretisation of the Kohn-Sham generalised eigenvalue problem in Eqn.~\eqref{eq: PAW-GHEP} results in an algebraic generalized Hermitian eigenvalue problem $\bH^{\bk,\sigma}\feNodalVectorUtilde^\sigma_{i,\bk} = \varepsilon^{\sigma,h}_{i,\bk}\bS^{\bk}\feNodalVectorUtilde^\sigma_{i,\bk}$, where $\bH^{\bk,\sigma}$ and $\bS^{\bk}$ denote the discretized Hamiltonian matrix and the PAW overlap matrices of size $M \times M$ with $\varepsilon^{\sigma,h}_{i,\bk}$ denoting the $i$\textsuperscript{th} eigenvalue corresponding to the discrete eigenvector 
$\feNodalVectorUtilde^\sigma_{i,\bk}$. The derivation of the above FE discretised generalised eigenvalue problem follows in a straightforward manner from the spin-unpolarized case discussed in Appendix C of our prior work~\cite{pawfe}. We note that the FE-discretized Hamiltonian matrix $\bH^{k,\sigma} = \bH^{\bk,\sigma,\text{loc}} + \bH^{\bk,\sigma,\text{nloc}}$ and FE-discretized PAW overlap matrix $\bS^{k} = \bM + \bS^{\bk,\text{nloc}}$ has both local and non-local contributions, where the entries of the local contribution are:

\begin{align}
\text{H}_{IJ}^{\bk,\sigma,\text{loc}} &= \bigintssss_{\Omega_p}{\Biggl\{ \frac{1}{2}\nabla N^{h,p}_I(\bx)\cdot \nabla N^{h,p}_J(\bx)}{+\;\left[V^{\sigma,h}_{\text{eff}}(\bx) + V_z(\bx) \right] N_I^{h,p}(\bx)N_J^{h,p}(\bx)}{+\; \frac{1}{2}|\bk|^2N_I^{h,p}(\bx)N_J^{h,p}(\bx) }\nonumber \\
& {- \mathrm{i}\bk \cdot N^{h,p}_I(\bx)\nabla N^{h,p}_J(\bx)+\bV^h_{\text{GGA}}(\bx) \cdot \left(N_I^{h,p}(\bx)\nabla N^{h,p}_J(\bx)+N^{h,p}_J(\bx)\nabla N^{h,p}_I(\bx) \right)\Biggr\} d\bx} \label{eqn:hij} \\
 \text{M}_{IJ} &= \int_{\Omega_p}{N_I^{h,p}(\bx)N_J^{h,p}(\bx)d\bx}  \label{eqn:mij} \\
 \text{with}\;\;&V^{\sigma,h}_{\text{eff}} =  \left[V_{\text{el}}^{h,p_{\text{el}}}(\bx) +\frac{\partial \epsilon_{\text{xc}}(\nTilde^\sigma, \nabla \nTilde^\sigma)}{\partial \nTilde^{\sigma}}\bigg\rvert_{\nTilde^{\sigma} = \nTilde^{\sigma,h}}  \right],\;\; \bV^{\sigma,h}_{\text{GGA}}(\bx) = \frac{\partial \epsilon_{\text{xc}}(\nTilde^{\sigma}, \nabla \nTilde^{\sigma})}{\partial \nabla \nTilde^{\sigma}}\bigg\rvert_{\nTilde^{\sigma} = \nTilde^{\sigma,h}} \nonumber
\end{align}
$V_{\text{el}}^{h,p_{\text{el}}}(\bx)$ in the above equation is determined as the solution of FE-discretised Poisson problem
\begin{equation}
    \frac{1}{4\pi}\bigintssss_{\Omega_p}\left[\nabla N^{h,p_{\text{el}}}_I(\bx)\cdot \nabla N^{h,p_{\text{el}}}_J(\bx) d\bx\right] V^{J}_{\text{el}} = \bigintssss_{\Omega_p}\left[N^{h,p_{\text{el}}}_I(\bx)\rhoTilde^h(\bx)\right]d\bx
    \label{eq: FE discretized Poisson problem}
\end{equation}
with appropriate boundary conditions applied. Finally, the matrix entries corresponding to the non-local contributions in $\bH^{\bk}$ and $\bS^{\bk}$ are given by
\begin{equation}
\bH^{\bk,\sigma,\text{nloc}} = \sum_a{\bC^{a,\bk}\boldsymbol{\Delta}^{a,\sigma}_{\bH}\bC^{a,\bk^{\dagger}}},\; \bS^{\bk,\text{nloc}} = \sum_a{\bC^{a,\bk}\boldsymbol{\Delta}^a_{\bS} \bC^{a,\bk^{\dagger}}}\label{eq: FE nonlocal contribution}
\end{equation}
with the matrix entries of $\bC^{a,\bk}$ defined as 
\begin{equation}
\text{C}^{a,\bk}_{J \alpha} = \int_{\Omega_p}\sum_r{e^{-\mathrm{i}\bk\cdot(\bx-\bL_r)}\pTilde^a_{\alpha}(\bx-(\bR^a+\bL_r))N^{h,p}_J(\bx)d\bx}
\end{equation}
where the atom-dependent coupling matrix $\boldsymbol{\Delta}^a_{\bH}$,  $\boldsymbol{\Delta}^a_{\bS}$ are of size $n^{a}_{\text{pj}}\cross n^{a}_{\text{pj}}$ with $n^{a}_{\text{pj}}$ denoting the number of projectors for atom index $a$. The corresponding matrix entries of $\boldsymbol{\Delta}^a_{\bH}$ are denoted as $\Delta h^{a,h}_{\alpha\beta}$ given by
\begin{align}
    & \Delta h^{a,\sigma,h}_{\alpha\beta} = \Delta T^a_{\alpha\beta} + \Delta C^a_{\alpha\beta} + 2\sum_{\alpha'\beta'}\Delta C^a_{\alpha\beta\alpha'\beta'}D^{a,\text{tot},h}_{\alpha'\beta'} + \Delta v^{a,\sigma}_{\text{xc},\alpha\beta}  \nonumber \\
  &+\sum_{lm}{\Delta^a_{lm\alpha\beta}\int_{\Omega_a}{\gTilde^a_{lm}(\bx)V^{h,p_{\text{el}}}_{\text{el}}(\bx)d\bx}} -\int_{\Omega_a}v^a_{\text{z}}(\bx)\phips^a_{\alpha}(\bx)\phips^a_{\beta}(\bx)d\bx  \label{eqn:deltaHanddeltaS}
\end{align}
while the matrix entries corresponding to $\boldsymbol{\Delta}^a_{\bS}$ are given by $\Delta s_{\alpha \beta}^a$ defined in Eqn. S-17 of SI.
Finally, the FE-discretised governing equations that are to be solved to determine the ground-state solution are
\begin{gather}
\bH^{\bk,\sigma}\feNodalVectorUtilde^\sigma_{i,\bk} = \varepsilon^{\sigma,h}_{i,\bk}\bS^{\bk}\feNodalVectorUtilde^{\sigma}_{i,\bk}\;\; \forall \bk \in BZ,\;\; \forall \sigma \in \{\uparrow, \downarrow \} \nonumber \\
 \sum_{\sigma}\sum_i^N{\fint_{BZ}f(\varepsilon^{\sigma,h}_{i,\bk},\mu)d\bk} = N_v,\; f(\varepsilon^{\sigma,h}_{i,\bk},\mu) = \frac{1}{1+ \text{exp}\left(\frac{\varepsilon^{\sigma,h}_{i,\bk}-\mu}{k_{\text{B}}T} \right)} \nonumber \\
\nTilde^h(\bx) = \nTilde^h_c(\bx) + \sum_{\sigma}\sum_i^N{\fint_{BZ}f(\varepsilon^{\sigma,h}_{i,\bk},\mu)|\fePsips^{\sigma,h}_{i,\bk}(\bx)|^2d\bk},\;\mTilde^h(\bx) = \sum_i^N{\fint_{BZ}\left[f(\varepsilon^{\uparrow,h}_{i,\bk},\mu)|\fePsips^{\uparrow,h}_{i,\bk}(\bx)|^2-f(\varepsilon^{\downarrow,h}_{i,\bk},\mu)|\fePsips^{\downarrow,h}_{i,\bk}(\bx)|^2\right]d\bk}  \nonumber \\
\widehat{\boldsymbol{\mathsf{U}}}^{a,\sigma,\bk}_{i} = \bC^{a,\bk\dagger}\feNodalVectorUtilde^{\sigma}_{i,\bk},\nonumber \\
D^{a,\sigma,h}_{\alpha\beta} = \sum_i^N{\fint_{BZ}{f(\varepsilon^{\sigma,h}_{i,\bk},\mu)\widehat{{\mathsf{U}}}^{a,\sigma,\bk^*}_{\alpha i}\widehat{{\mathsf{U}}}^{a,\sigma,\bk}_{\beta i} d\bk}},\;D^{a,\text{mag},h}_{\alpha\beta}:= \left( D^{a,\uparrow,h}_{\alpha\beta} -D^{a,\downarrow,h}_{\alpha\beta}\right), \;\;D^{a,\text{tot},h}_{\alpha\beta}:= \left( D^{a,\uparrow,h}_{\alpha\beta} + D^{a,\downarrow,h}_{\alpha\beta}\right) \nonumber \\
\bTilde^{a,h}(\bx) = \kappa^a\gTilde^a_0(\bx) + \sum_{lm,\alpha\beta}{\Delta^a_{lm\alpha\beta}D^{a,\text{tot},h}_{\alpha\beta}\gTilde^a_{lm}(\bx)},\; \bTilde^h(\bx) = \sum_a{\bTilde^{a,h}(\bx)},\; \rhoTilde^h(\bx) = \nTilde^h(\bx)+\bTilde^h(\bx) \nonumber
\label{eq: FE PAW governing equations}
\end{gather}
where $\bC^{a,\bk}$ is defined in Eqn.~\eqref{eq: FE nonlocal contribution} and the electrostatic potential $V^h_{\text{el}}$ is determined from Eqn.~\eqref{eq: FE discretized Poisson problem}.
Using the solution of Eqn.~\eqref{eq: FE PAW governing equations}, the ionic forces, all-electron density, and all-electron wavefunctions are evaluated using Eqns.~\eqref{eq: ionic forces PAW}, \eqref{eq: AE-density}, and \eqref{eq: AE-wfc}, respectively.

%% file: computationalMethodology.tex
\cb 
 Building on the foundation presented in the previous section, this section presents the primary algorithmic contribution of this work: a residual-based reformulation of the Chebyshev-filtered subspace iteration (R-ChFSI) that enables the efficient solution of the large, sparse PAW generalized eigenproblem arising at each SCF iteration. \cn

\subsection{Residual-based Chebyshev filtered subspace iteration for PAW eigenproblem}
As commonly employed in most real-space codes~\cite{sparc2017a, PARSEC, dftfe1.0}, the Chebyshev filtered subspace iteration~\cite{ChFSISAAD2006, SaadChFSIBook} is the method of choice to solve the nonlinear eigenvalue problem encountered in DFT. However, the method was initially employed to solve standard eigenvalue problem and subsequently was extended to solve the PAW generalised eigenvalue problem~\cite{LEVITT201598, pawfe}. In this method, we first construct a Chebyshev-filtered (CF) subspace by applying a polynomial filter of $\bS^{-1}\bH$ to an initial trial subspace, followed by Rayleigh-Ritz projection (RR-Projection) and subspace diagonalisation with subspace rotation (RR-GHEP and RR-SR). Specifically, the filtered subspace is obtained as $\Psips_{\text{f}} = T_m(\widetilde{\bH})\Psips_{\text{in}}$ where $T_m(\widetilde{\bH})$ denotes the $m^{th}$ degree Chebyshev polynomial of the matrix $\widetilde{\bH}$ and the columns of the $M\times N$ matrix $\Psips_{\text{in}}$ represent approximations to the desired eigenvectors of ${\bS}^{-1}\bH$ that has the same eigenspace of $\widetilde{\bH} = \frac{2}{b_U-b_W}\left( {\bS}^{-1}\bH\right) -\frac{b_U+b_W}{b_U-b_W}\boldsymbol{I}$. The parameters $b_U$ and $b_W$ are chosen such that target eigenvalues of $\mathbf{S}^{-1}\mathbf{H}$ are mapped to values strictly less than -1, while the unwanted eigenvalues are mapped into the interval $[-1,1]$. This spectral transformation enables the Chebyshev polynomial to selectively amplify components associated with the desired eigenspace while suppressing the rest. To ensure that the amplification of the desired eigenspace does not lead to floating-point overflow, we instead build the Chebyshev filtered space as $\Psips_{\text{f}} = \mathcal{C}_m(\widetilde{\bH}, \widetilde{a_0}) := {T_m(\widetilde{\bH})}/{T_m(\widetilde{a}_0)}$, where 
$a_0$ is an estimate of the lowest eigenvalue of $\bS^{-1}\bH$ and $\widetilde{a}_0$ is scaled and shifted $a_0$ using the same affine transformation that maps  $\bS^{-1}\bH$ to $\widetilde{\bH}$. 
Consequently, the Chebyshev filtered space is constructed  using the three-term recurrence relation of Chebyshev polynomials  as given below:
\begin{align}
&\Psips_\text{f} = \mathcal{C}_m(\widetilde{\bH},\widetilde{a}_0)\Psips_{\text{in}} \;\text{where}\; \mathcal{C}_m(\widetilde{\bH},\widetilde{a}_0) = T_m(\widetilde{\bH})/ T_m(\widetilde{a}_0) \nonumber \\
&\text{and}\; T_{\theta+1}(\bX) = 2\bX T_{\theta}(\bX) - T_{\theta-1}(\bX),\;\; \text{for}\; 1 \leq \theta \leq m-1 \nonumber \\
&\text{with}\;\; T_0(\bX) = \bI, T_1(\bX) = \bX
   \label{eq: ChFSI recurrance}
\end{align}

Using Eqn.~\eqref{eq: ChFSI recurrance}, the Chebyshev filtered space is constructed using the following recursive relation involving matrix-matrix products
\begin{equation}
    \bX_{\theta+1} = \frac{2\zeta_{\theta+1}}{e}\left( {\bS}^{-1}\bH\right)\bX_{\theta} -\frac{2\zeta_{\theta+1}c}{e}\bX_{\theta} - \zeta_{\theta}\zeta_{\theta+1}\bX_{\theta-1},\;\; \text{for}\; 2 \leq \theta \leq m 
    \label{eq: ChFSI recurrance2}
\end{equation}
where $\bX_0 = \Psips_{\text{in}}$, $\bX_1 = \frac{\zeta_1}{e}(\left( {\bS}^{-1}\bH\right)-c\bI)\Psips_{\text{in}}$ and $\zeta_{\theta+1} = \frac{1}{\left(\frac{2}{\zeta_1} - \zeta_{\theta}\right)}$ with $\zeta_1 = \frac{e}{\left(a_0-c\right)}$, $e = \frac{\left(b_U-b_W\right)}{2}$ and $c = \frac{\left(b_W+b_U\right)}{2}$.
The rapid growth property of Chebyshev polynomials outside $[-1,1]$ amplifies the components in $\Psips_{\text{in}}$ along the direction of the desired eigenvectors (occupied states) while dampening the components along the unwanted eigenvectors (unoccupied states), leading to the filtered subspace $\Psips_{\text{f}}$ being a close approximation of the eigenspace of interest of the PAW eigenproblem $\bH\bfePsips=\varepsilon \bS\bfePsips$ in a given SCF iteration. The values of $b_U$, $a_0$ correspond to the upper bound of the unwanted spectrum and the lower bound of the wanted spectrum, which are both estimated from a few iterations of the Generalised Lanczos iteration method as discussed in our previous work~\cite{pawfe}. Furthermore, the value of $b_W$ corresponding to the upper bound of the wanted spectrum is estimated from the eigenvalues of the previous SCF iteration or from $a_0$ and $b_U$.

\subsubsection{Proposed filtering approach leveraging approximate inverse of PAW overlap matrix:} In this work, instead of using the recurrence relation Eqn.~\eqref{eq: ChFSI recurrance2} to construct the filtered space $\Psips_{\text{f}}$, we formulate this three-term recurrence relation in terms of the eigenproblem residuals [$\bH\Psips_{\text{in}}-\bS\Psips_{\text{in}}\boldsymbol{\Lambda}_{\text{in}}$] in the spirit of our recent work R-ChFSI~\cite{RChFSI}. Here, we note that the $\boldsymbol{\Lambda}_{\text{in}}$ is the diagonal matrix whose entries are an estimate for the eigenvalues of the wanted eigenspace. In contrast to the traditional Chebyshev filtered subspace iteration approach, we show that the residual-based reformulation is significantly more robust to approximations in matrix–multivector products during the construction of the filtered subspace for the PAW-GHEP. As demonstrated later, this robustness enables a more efficient evaluation of the action of $\bS^{-1}$ on GPUs by permitting the use of inexpensive approximate inverses. Furthermore, it enables the efficient construction of the filtered subspace using low-precision compute and communication as discussed in the subsequent section.
 
To derive the recurrence relation in terms of residuals, we first introduce
\begin{align}
    \bZ_{\theta} = \bS \mathcal{C}_{\theta}(\widetilde{\bH},\widetilde{a}_0)\Psips_{\text{in}} = \bS\bX_{\theta} ,\;\;
    \bL_{\theta} = \bS\Psips_{\text{in}}\mathcal{C}_{\theta}(\widetilde{\boldsymbol{\Lambda}}_\text{in},\widetilde{a}_0)
\end{align}
where $\widetilde{\boldsymbol{\Lambda}}_\text{in}$ is the affine transformation of $\boldsymbol{\Lambda}_{\text{in}}$ using the same affine map that transforms $\bS^{-1}\bH$ to $\widetilde{\bH}$ and $\mathcal{C}_{\theta}(\widetilde{\boldsymbol{\Lambda}}_\text{in},\widetilde{a}_0) = T_{\theta}(\widetilde{\boldsymbol{\Lambda}}_\text{in})/T_{\theta}(\widetilde{a}_0)$ which can be expressed using the recurrence relation mentioned in Eqn.~\eqref{eq: ChFSI recurrance}. To this end, the recurrence relation for $\bZ_{\theta}, \bL_{\theta}$  similar to Eqn.~\eqref{eq: ChFSI recurrance2} can be expressed as
\begin{align}
     & \bZ_{\theta+1} = \frac{2\zeta_{\theta+1}}{e}\left( \bH{\bS}^{-1}\right)\bZ_{\theta} -\frac{2\zeta_{\theta+1}c}{e}\bZ_{\theta} - \zeta_{\theta}\zeta_{\theta+1}\bZ_{\theta-1} \\
  &    \bL_{\theta+1} = \frac{2\zeta_{\theta+1}}{e}\bL_{\theta}\left( \boldsymbol{\Lambda}_{\text{in}}\right) -\frac{2\zeta_{\theta+1}c}{e}\bL_{\theta} - \zeta_{\theta}\zeta_{\theta+1}\bL_{\theta-1}  
  \label{eqn: expression for Z and L}
\end{align}
with $\bZ_0 = \bS\Psips_{\text{in}}, \bZ_1 = \frac{\zeta_1}{e}(\bH - c\bS)\Psips_{\text{in}}, \bL_0 = \bS\Psips_{\text{in}}, \bL_1 = \frac{\zeta_1}{e}\bS\Psips_{\text{in}}(\boldsymbol{\Lambda}_{\text{in}} - c\bI)$.

We  now define $\bR_\theta$ as $\bR_\theta =  \bZ_{\theta} - \bL_{\theta} $, which evaluated
 at $\theta= \{0,1\}$ gives:
 \begin{equation}
 \bR_0 = \boldsymbol{0}, \bR_1 = \frac{\zeta_1}{e}\left[\bH\Psips_{\text{in}}-\bS\Psips_{\text{in}}\boldsymbol{\Lambda}_\text{in}\right]
 \end{equation}
 and for $2\leq\theta\leq m$
 we derive a recurrence relation on $\bR_{\theta}$ as:
\begin{equation}
    \bR_{\theta+1} = \frac{2\zeta_{\theta+1}}{e}\bH{\bS}^{-1}\bR_{\theta} -\frac{2\zeta_{\theta+1}c}{e}\bR_{\theta} - \zeta_{\theta}\zeta_{\theta+1}\bR_{\theta-1} + \frac{2\zeta_{\theta+1}}{e}\left(\bH\Psips_{\text{in}}-\bS\Psips_{\text{in}}\boldsymbol{\Lambda}_\text{in}\right)\boldsymbol{\Lambda}_{\theta} \label{eq: RChFSI residual 0}
    \end{equation}
    \begin{equation}
  \text{where}\;  \boldsymbol{\Lambda}_{\theta+1} = \frac{2\zeta_{\theta+1}}{e}\boldsymbol{\Lambda}_{\text{in}}\boldsymbol{\Lambda}_{\theta} -\frac{2\zeta_{\theta+1}c}{e}\boldsymbol{\Lambda}_{\theta} - \zeta_{\theta}\zeta_{\theta+1}\boldsymbol{\Lambda}_{\theta-1}
    \label{eq: RChFSI residual 1}
\end{equation}
Finally, the filtered space $\Psips_{\text{f}}$ is recovered from Eqn.~\eqref{eq: RChFSI residual 0} at $\theta=m$ as: 
\begin{equation}
  \Psips_\text{f}  = {\bS}^{-1}\bR_m + \Psips_{\text{in}}\mathcal{C}_m(\widetilde{\boldsymbol{\Lambda}}_{\text{in}},\widetilde{a}_0)  
  \label{eq: RChFSI final}
\end{equation}

We note here that the recurrence relation derived here is mathematically equivalent to Eqn.~\eqref{eq: ChFSI recurrance} but the recurrence relation now is expressed in-terms of eigenproblem residuals. The motivation behind this choice of building the filtered space lies in the fact that both $\mathcal{C}_m(\widetilde{\bH},\widetilde{a}_0)$ and $\left( {\bS}^{-1}\bH\right)$ share the same wanted eigenspace while their eigenvalues are $\mathcal{C}_m(\widetilde{\boldsymbol{\Lambda}},\widetilde{a}_0)$ and $\boldsymbol{\Lambda}$ respectively. Consequently, as the estimates of the desired eigenspace $\Psips_{\text{in}}$ and eigenvalues $\boldsymbol{\Lambda}_{\text{in}}$ improve across successive iterations, the residual $\bR_1 = \bH\Psips_{\text{in}}-\bS\Psips_{\text{in}}\boldsymbol{\Lambda}_\text{in}$ tends to zero. Consequently, $\bR_{\theta+1}$ for $1 \le \theta \le m$ also tends to $\boldsymbol{0}$, and hence the effect of approximations introduced in evaluating the action of $\bS^{-1}$ or $\bH$ on $\bR_\theta$ diminishes. 
Now, we exploit this property of the new recurrence relation in Eqns.~\eqref{eq: RChFSI residual 0},~\eqref{eq: RChFSI residual 1} to introduce approximations that enable a computationally tractable evaluation of $\bS^{-1}$ as discussed below.  Since $\bS$ is a sparse matrix of dimension $M \times M$, explicitly computing its inverse is computationally prohibitive. Therefore, as was described in our previous work~\cite{pawfe} we compute $\widehat{\bS}^{-1}$, a close approximation to $\bS^{-1}$ which is given by
\begin{align}\label{eqn:Sinvapprox}
\widehat{\bS}^{-1} &= \bM_{\text{D}}^{-1} - \sum_a{{\bM_{\text{D}}^{-1}}\bC^a{\boldsymbol{\Delta}^a _{\bS_{\text{inv}}}}\bC^{a\dagger}} \bM_{\text{D}}^{-1} \\
\text{with}\;\; \boldsymbol{\Delta}^a _{\bS_{\text{inv}}}&= {\left({\boldsymbol{\Delta}_{\bS}^{a^{-1}}} + {\bC^{a}}^{\dagger} \bM_{\text{D}}^{-1}\bC^{a} \right)}^{-1} \label{eq: Delta Sinv}
\end{align}
\cb where we introduced two approximations in the above two equations: (i) using the  Gauss-Lobatto-Legendre quadrature, a reduced order quadrature rule with quadrature points coincident with the FE-nodal points, we approximate the FE-overlap matrix to be diagonal ($\bM_\text{D}$) (ii) exploiting the fact that PAW augmentation spheres seldom overlap, the overlap between projector functions associated with different atoms is neglected, resulting in a block-diagonal approximation of the overlap matrix corresponding to projector functions (refer to $\bP$ matrix defined in Eqn.~37 of our previous work~\cite{pawfe}), with each block corresponding to an individual atom. We emphasize here, that this approximation is possible only in real-space as the augmentation spheres have compact support in real-space unlike in PW methods where these operators are evaluated in Fourier space~\cite{LEVITT201598}. \cn
In Eqn.~\eqref{eqn:Sinvapprox}, $\bM_{\text{D}}$ is as defined above. Additionally, to reduce the cost of evaluating $\boldsymbol{\Delta}^a _{\bS_{\text{inv}}}$, especially for large system sizes on limited computational resources, we re-examine  Eqn.~\eqref{eq: Delta Sinv}, which is required to evaluate $\widehat{\bS}^{-1}$.
Upon closer inspection, we observe that $(\bC^{a \dagger} \bM^{-1}\bC^{a})_{\alpha\beta} = \sum_{IJ}\braket{\pTilde^a_{\alpha}|N^I}M^{-1}_{IJ}\braket{N^J|\pTilde^a_{\beta}}$ can be approximated as $\braket{\pTilde^a_{\alpha}|\pTilde^a_{\beta}}$ by taking cognizance of the fact that $\sum_{IJ}\ket{N^I}M^{-1}_{IJ}\bra{N^J} \rightarrow \boldsymbol{\mathbb{I}}$, the identity operator in the limit of FE-mesh refinement. Now, Eqn.~\eqref{eq: RChFSI residual 0} can be expressed in terms of $\widehat{\bS}^{-1}$ as
\begin{equation}
 \bR_{\theta+1} = \frac{2\zeta_{\theta+1}}{e}\bH\widehat{\bS}^{-1}\bR_{\theta} -\frac{2\zeta_{\theta+1}c}{e}\bR_{\theta} - \zeta_{\theta}\zeta_{\theta+1}\bR_{\theta-1} + \frac{2\zeta_{\theta+1}}{e}\left(\bH\Psips_{\text{in}}-\bS\Psips_{\text{in}}\boldsymbol{\Lambda}_\text{in}\right)\boldsymbol{\Lambda}_{\theta} \label{eq: RChFSI residualfinal 0}   
\end{equation}
and finally the filtered space $\Psips_{\text{f}}$ is computed as 
$
  \Psips_\text{f}   = \widehat{\bS}^{-1}\bR_m + \Psips_{\text{in}}\mathcal{C}_m(\widetilde{\boldsymbol{\Lambda}}_{\text{in}},\widetilde{a}_0) $

 Finally, the various steps involved in solving the PAW-GHEP are outlined in Algo~\eqref{alg: ChFSI}, where Step 3 corresponds to the Chebyshev filtered space construction using the residual-based approach that is summarised in Algo~\eqref{alg: CF: Residual Cheby} and Steps 4-6 correspond to the Rayleigh-Ritz step as is usually done in iterative orthogonal projection based eigensolvers.
\begin{figure}
\begin{algorithm}[H]
    \caption{Chebyshev Filtered subspace iteration algorithm in \pawfe}\label{alg:ChFSI}
\KwIn{Initial subspace $\Psips_{\text{in}}$}
\KwOut{$\bS$-orthonormalised filtered subspace $\Psips$}
\begin{enumerate}
\item Estimate $[b_0,b_U]$ from k-step Generalised Lanczos (see Algo.3 in the appendix of our previous work~\cite{pawfe}). '$b_W$' can be estimated from $[b_0,b_U]$ or estimated from the previous SCF iteration.
\item Scale and shift the Hamiltonian: $\widetilde{\bH} = \frac{2}{b_U-b_W}\left( \widehat{\bS}^{-1}\bH\right) -\frac{b_W+b_U}{b_U-b_W}\boldsymbol{I}$
\item $[$\texttt{CF}$]$ Compute Chebyshev filtered subspace:    $\Psips_\text{f} = T_m(\widetilde{\bH})\Psips_{\text{in}}$. ($\mathcal{O}(MN)$)
\item $[$\texttt{RR-Projection}$]$ Compute the projected Hamiltonian ($\bH^{p}$) and projected PAW overlap matrices ($\bS^{p}$): $\bH^{p} =\Psips_{\text{f}}^{\dagger}\bH\Psips_{\text{f}} $, $\bS^{p} = \Psips_{\text{f}}^{\dagger}\bS\Psips_{\text{f}}$ ($\mathcal{O}(MN^2)$)
\item $[$\texttt{RR-GHEP}$]$ Solve the GHEP: $\bH^{p}\bQ = \bS^{p}\bQ\boldsymbol{\Lambda}$ ($\mathcal{O}(N^3)$)
\item $[$\texttt{RR-SR}$]$ Perform subspace rotation:$\Psips = \Psips_{\text{f}}\bQ$. ($\mathcal{O}(MN^2)$)
\end{enumerate}
\label{alg: ChFSI}
\end{algorithm}
\end{figure}

\begin{algorithm}[H]
    \caption{Construction of  Residual based Chebyshev filtered subspace} \label{alg: CF: Residual Cheby}
\KwIn{(i) Chebyshev polynomial order $m$, (ii)estimates of the bounds of the eigenspectrum $[b_U,b_0]$, (iii) estimate of the upper bound of the wanted spectrum $b_W$, (iv)the initial guess of eigenpairs   ($\{\boldsymbol{\Lambda},\Psips_{\text{in}}\}$)}
\KwOut{The filtered space of eigenvector block  ($\Psips_{\text{f}}$)} 
\begin{enumerate}
\item {$e \leftarrow \frac{b_U-b_W}{2}$; $c \leftarrow \frac{b_U+b_W}{2}$; $\zeta \leftarrow \frac{e}{b_0-c}$; $\zeta_1 \leftarrow \zeta$; $\gamma \leftarrow \frac{2}{\zeta_1}$}
\item{ $\bX = \Psips_{\text{in}}$,$\bY \leftarrow \bH\bX - \bS\bX\boldsymbol{\Lambda}$}
\item {$\bR_{\bX} \leftarrow  0$, $\bR_{\bY} \leftarrow \frac{\zeta_1}{e}\bY$}, $\boldsymbol{\Lambda}_{\bX}\leftarrow \bI$, $\boldsymbol{\Lambda}_{\bY}\leftarrow \frac{\zeta_1}{e}\left(\boldsymbol{\Lambda} -c\bI  \right)$
\item for $\theta = 2$ to $m$ do 
\begin{itemize}
    \item {$\zeta_1\leftarrow \frac{1}{\gamma-\zeta}$}
    \item {$\bR_{\bX}\leftarrow\frac{2\zeta_1}{e}\left(\bH\widehat{\bS}^{-1}\right)\bR_{\bY}-\frac{2\zeta_1}{e}c\bR_{\bY}-\zeta\zeta_1\bR_{\bX} + \frac{2\zeta_1}{e}\bY\boldsymbol{\Lambda}_{\bY}$}
    \item $\boldsymbol{\Lambda}_{\bX}  \leftarrow\frac{2\zeta_1}{e}\boldsymbol{\Lambda}_{\bY}\boldsymbol{\Lambda} - \frac{2\zeta_1}{e}c\boldsymbol{\Lambda}_{\bY} -\zeta\zeta_1\boldsymbol{\Lambda}_{\bX}  $
    \item {$\left(\bR_{\bX}\leftrightarrow\bR_{\bY}\right)$},{$\left(\boldsymbol{\Lambda}_{\bX}\leftrightarrow\boldsymbol{\Lambda}_{\bY}\right)$ and  $\zeta = \zeta_1$}
\end{itemize}
\item return $\left(\widehat{\bS}^{-1}\bR_{\bY} + \bX\boldsymbol{\Lambda}_{\bY}\right)$
\end{enumerate}
\end{algorithm}

%% file: NumericalImplementation.tex
In this section, we propose various algorithmic strategies and implementation innovations to determine the ground-state DFT solution efficiently using the computational methods discussed in the previous section on modern GPU architectures.  To this end, this section highlights the synergy between the projector-augmented-wave (PAW) formalism and the inherent advantages of the finite element method that leverage extreme levels of parallelism offered by evolving computing architectures. As discussed previously, the PAW generalised eigenproblem (PAW-GHEP) $\bH\Psips = \bS\Psips\boldsymbol{\Lambda}$ is solved in each self-consistent field (SCF) iteration using the Chebyshev filtered subspace iteration approach (ChFSI).  In particular, we resort to the residual-based reformulation of the Chebyshev filtered subspace iteration approach (R-ChFSI)~\cite{RChFSI}, tolerant to approximations in matrix-multivector multiplications as discussed previously. Recalling from section~\ref{sec: 3}, the FE discretised operators $\bH$ and $\bS$ are $M\times M$ sparse matrices and 
$\Psips$ denotes the $M \times N$ multi-vector comprising FE-discretized pseudo-wavefunctions
where '$M$' denotes the total degrees-of-freedom (DoFs) to be solved and '$N$' denotes the number of wavefunctions.

Furthermore, as illustrated in Algorithm~\ref{alg:ChFSI}, the Chebyshev filtered space construction scales quadratically with system size and requires the action of  
$\bH\widehat{\bS}^{-1}$ on multivectors. Following this, we perform the Rayleigh-Ritz step, which scales cubically with system size and involves: (i) projecting the FE-discretised PAW GHEP onto space spanned by the columns of filtered space ($\Psips_{\text{f}}$), (ii) solving the smaller projected generalized eigenvalue problem, and (iii) performing a subspace-rotation step to obtain the \textbf{S}-orthonormal eigenvectors ($\Psips$). 

At the end of each R-ChFSI iteration, the eigenproblem residual matrix $\bR = \left(\bH\Psips - \bS\Psips\boldsymbol{\Lambda}\right)$ is computed. We ensure that $||\br_i||_2 \leq 1\times 10^{-2}$ at each SCF iteration, where $\br_i$ refers to the $i^{th}$ column of  $\bR$ and $1\leq i \leq N_f$ with $N_f$ the index of the lowest eigenstate above the Fermi energy. If the condition $||\br_i||_2 \leq \delta_{\text{res}}$ is not satisfied, we perform multiple iterations of R-ChFSI using $\Psips$ and $\boldsymbol{\Lambda}$ from the previous iteration as initial guesses for subsequent iteration.  We note that as the SCF iterations proceed, the residuals $||\br_i||_2$ for $1\leq i \leq N_f$  satisfy $\delta_{\text{res}}$ in a single iteration of R-ChFSI for a given SCF and the final values of $||\br_i||_2$ are around $\mathcal{O}(10^{-7})-\mathcal{O}(10^{-9})$ for $1\leq i \leq N_f$ at SCF convergence.

Consequently, in each SCF iteration, we require efficient strategies for evaluating the FE-discretised matrices, computing the action of the various FE-discretised PAW operators on multivectors involved in \texttt{CF} step, evaluating the products of the form $\bX^{\dagger}\bY$ in a distributed setting, solving the projected eigenproblem in \texttt{RR-GHEP} step, and carrying out the subspace rotation step in \texttt{RR-SR}. We detail the efficient numerical implementation of each of these steps in the remaining part of this section.
\subsection{Evaluating the FE-Discretized Matrices}~\label{sec: 4.1}
Denoting the simulation domain as $\Omega_p$, we begin by partitioning $\Omega_p$ into  subdomains $\Omega^{(t)}\,\forall t = 1,2,...,n_t$, where $n_t$ represents the number of subdomains, with each subdomain $\Omega^{(t)}$ assigned to an \texttt{MPI-task} `$t$'. Furthermore, each $\Omega^{(t)}$ is further subdivided into $E_t$ non-overlapping subdomains represented as $\Omega^{(e,t)}$ such that $\Omega^{(t)} =  \cup^{E_t}_{e=1}\Omega^{(e,t)}$, where the notation $(e,t)$ refers to index of the FE-cell `$e$' present in MPI task `$t$'. This allows the integrals to be decomposed over the simulation domain $\Omega_p$ in Eqns.~\eqref{eqn:hij}, \eqref{eq: FE nonlocal contribution}, \eqref{eq: FE discretized Poisson problem} into a series of integrals over smaller domains $\Omega^{(e,t)}$ as $\int_{\Omega_p}f(\bx)d\bx = \sum_t\sum_e^{E_t}\int_{\Omega^{(e,t)}}f(\bx)d\bx$.  
Subsequently, the integrals over each FE-cell domain $\Omega^{(e,t)}$ are computed using a suitable quadrature rule as:
\begin{align}
    \int_{\Omega^{(e,t)}}{f(\bx)d\bx} =& \int_{\widehat{\Omega}}{f(\widehat{\bx})\det \{\bJ_{(e,t)}\}d\widehat{\bx}} \nonumber \\  
    \approx & \cn \sum^{n_q}_q{w_qf(\bx_q)\det \{\bJ_{(e,t)}\}\biggr\rvert_{\widehat{\bx}_q}}
    \label{eq: Quadrature Integration}
\end{align}
where $\widehat{\Omega}=[-1,1]^3$ refers to the reference cell, $n_q$ is the total number of quadrature points in the domain $\widehat{\Omega}$. Furthermore,  $\widehat{\bx}\in \widehat{\Omega}$ maps to a point in the discretised cell $\bx \in \Omega^{(e,t)}$ and the $3\times 3$ matrix $\bJ_{(e,t)}$ is the Jacobian of the mapping from the current FE-cell $\Omega^{(e,t)}$ to the reference cell $\widehat{\Omega}$ whose elements are ${J_{(e,t)}}_{ij} = \nicefrac{\partial x^{(e,t)}_i}{\partial \widehat{x}_j}\biggr\rvert_{\widehat{\bx}_q}$ as is usually done in finite-element based approaches \cite{hughes2012finite}.  3D Gauss-Legendre quadrature rules that are tensor products of 1D quadrature rules~\cite{glrules} are employed in this work, and the order of these rules employed is determined such that errors due to the quadrature rule are of higher order than the discretisation error. The flexibility in the choice of quadrature rules for evaluating various integrals involving PAW atomic data in  Eqn.~\eqref{eq: FE nonlocal contribution} allows the use of coarser finite-element meshes to represent electronic fields and yet capture the relevant atom-centered data terms. This is a notable advantage compared to the finite-difference approach, a popular real-space discretization DFT method, where special schemes need to be devised to retain the use of coarser grid resolution for smoother electronic fields without egg-box effects~\cite{parsecleo}. 

As introduced in Sec.~\ref{sec: 2}, $\sigma \in \{\uparrow, \downarrow\}$ denotes the spin index and $\bk$ represents a point in the first Brillouin zone, and are used to index the FE-discretized Hamiltonian $\bH^{\sigma,\bk}$ and PAW overlap matrices $\bS^{\bk}$. Furthermore, as discussed in the previous section, we 
employ $ n_p^3$-three-dimensional (3D) Lagrange polynomial basis functions that are tensor products of 1D-Lagrange polynomials of degree $p$, and $n_p = p+1$ denotes the number of points in each spatial direction of the FE-cell. These basis functions are constructed over GLL (Gauss-Lobatto-Legendre) nodal points~\cite{spectralGLL}. This choice ensures that the FE-basis function centred at nodal point '$K$' and denoted by $\{N^{h,p}_K\}_{1\leq K\leq M}$, has compact support and is non-zero only within those finite-element cells that share the nodal point $K$. As a consequence of this locality, the FE-discretised matrices $\bH^{\sigma,\bk,\text{loc}}, \bM$ are sparse and can be decomposed into cell-wise contributions ($\bH^{e,\sigma,\bk,\text{loc}},\bM^{e}$) of dimension $n_p^{3}\times n_{p}^{3}$. 
As is usually done in finite-element methods, we evaluate $\{N^{h,p}_I(\bx)\}_{1 \leq I \leq n_p^3} \; \forall \bx \in \Omega^{(e,t)}$ using the FEM shape functions constructed in the reference cell ($\widehat{\Omega}$) which is denoted as $\{\widehat{N}_I^{h,p}(\widehat{\bx})\}_{1\leq I \leq n^3_p}$ where $\widehat{\bx}\in \widehat{\Omega}$. Similarly, $\nabla N^{h,p}_I(\bx)$ is evaluated as $\nabla N^{h,p}_I(\bx) = \bJ^{-1}_{(e,t)}\biggr\rvert_{\widehat{\bx}}\nabla_{\widehat{\bx}}\widehat{N}_I^{h,p}(\widehat{\bx})$ where $\bJ$ is the Jacobian of the mapping discussed previously.
Finally, the cell-level $\bH^{e,\sigma,\bk,\text{loc}}$ ,$\bM^{e}$  and entries of the non-local projector matrix $C^{{ e,\bk,a}^\dagger}_{\alpha I}$ are given by
\begin{align}
H^{e,\sigma,\bk,\text{loc}}_{IJ} :=& \frac{1}{2}\sum_{q_1}^{n_{q_1}}{w_{q_1}\left(\bJ^{-1}_{(e,t)}\biggr\rvert_{\widehat{\bx}_{q_1}}\nabla_{\widehat{\bx}} \widehat{N}^{h,p}_I(\widehat{\bx}_{q_1})\right)\cdot \left( \bJ^{-1}_{(e,t)}\biggr\rvert_{\widehat{\bx}_{q_1}} \nabla_{\widehat{\bx}} \widehat{N}^{h,p}_J(\widehat{\bx}_{q_1})\right)\det\{\bJ_{(e,t)}\}\biggr\rvert_{\widehat{\bx}_{q_1}}}\nonumber \\
    & + \sum_{q_1}^{n_{q_1}}w_{q_1}\left[-i\bk\cdot\left(\bJ^{-1}_{(e,t)}\biggr\rvert_{\widehat{\bx}_{q_1}}\nabla_{\widehat{\bx}} \widehat{N}^{h,p}_I(\widehat{\bx}_{q_1})\right) \widehat{N}^{h,p}_J(\widehat{\bx}_{q_1}) \right. \nonumber \\
    & \left. +\frac{1}{2}|\bk|^2\widehat{N}^{h,p}_I(\widehat{\bx}_{q_1}) \widehat{N}^{h,p}_J(\widehat{\bx}_{q_1})\right]\det\{\bJ_{(e,t)}\}\biggr\rvert_{\widehat{\bx}_{q_1}}  \nonumber \\
    &+ \sum_{q_2}^{n_{q_2}}{w_{q_2}\widehat{N}^{h,p}_I(\widehat{\bx}_{q_2})\widehat{N}^{h,p}_J(\widehat{\bx}_{q_2})\left[V^{h,p_{\text{el}}}_{\text{el}}(\bx_{q_2})+V^{\sigma,h}_{\text{LDA}}(\bx_{q_2})+V_{\text{z}}(\bx_{q_2})\right]\det\{\bJ_{(e,t)}\}\biggr\rvert_{\widehat{\bx}_{q_2}}} \nonumber \\
    & + \sum_{q_2}^{n_{q_2}}w_{q_2}\bV^{\sigma,h}_{\text{GGA}}(\bx_{q_2})\cdot\left[\bJ^{-1}_{(e,t)}\biggr\rvert_{\widehat{\bx}_{q_2}}\nabla_{\widehat{\bx}} \widehat{N}^{h,p}_I(\widehat{\bx}_{q_2}) \widehat{N}^{h,p}_J(\widehat{\bx}_{q_2}) \right . \nonumber \\
&\left. +\widehat{N}^{h,p}_I(\widehat{\bx}_{q_2})\bJ^{-1}_{(e,t)}\biggr\rvert_{\widehat{\bx}_{q_2}}\nabla_{\widehat{\bx}} \widehat{N}^{h,p}_J(\widehat{\bx}_{q_2})\right]\det\{\bJ_{(e,t)}\}\biggr\rvert_{\widehat{\bx}_{q_2}} \label{eq: FEM Ham entries} \\
  M^{e}_{IJ}  :=& \sum_{q_1}^{n_{q_1}}{w_{q_1}\widehat{N}^{h,p}_I(\widehat{\bx}_{q_1}) \widehat{N}^{h,p}_J(\widehat{\bx}_{q_1})\det\{\bJ_{(e,t)}\}\biggr\rvert_{\widehat{\bx}_{q_1}}} \label{eq: FEM S_loc entries} \\
   C^{{ e,\bk,a}^\dagger}_{\alpha I} := & \sum_r\sum_{q_3}{w_{q_3}e^{i\bk\cdot(\bx_{q_3}-\bL_{\br})}\widehat{N}^{h,p}_I(\widehat{\bx}_{q_3})\pTilde^a_{\alpha}(\bx_{q_3}-\bR^a-\bL_{\br})\det\{\bJ_{(e,t)}\}\biggr\rvert_{\widehat{\bx}_{q_3}}} \label{eq: FEM Cmatrix}
\end{align}
Here, $V^{h,p_{\text{el}}}_{\text{el}}(\bx)$ is the FE-discretized electrostatic potential (as introduced in Sec.~\ref{sec: 2}), $V^{\sigma,h}_{\text{LDA}}(\bx)$ denotes the LDA-like scalar part of the exchange–correlation potential, and $\bV^{\sigma,h}_{\text{GGA}}(\bx)$ denotes the gradient correction to the exchange–correlation potential.
where the Gauss–Legendre quadrature rules are employed for numerical integration. The number of quadrature points for the kinetic-energy term, FE overlap term is chosen as $n_{q_1} = (p+1)^3$, which is sufficient to exactly integrate polynomials of degree up to $2p$ arising from evaluating the entries of finite-element overlap matrix ($M^{e}_{IJ}$). The quadrature orders $n_{q_2}$ and $n_{q_3}$ corresponding to the remaining terms are selected such that the associated quadrature error is at least one order of magnitude lower than the discretization error of the underlying finite-element mesh. The evaluation of the matrix entries in Eqs.~\eqref{eq: FEM Ham entries}, \eqref{eq: FEM S_loc entries}, and \eqref{eq: FEM Cmatrix} is reformulated as  matrix–matrix multiplications. These operations exhibit high arithmetic intensity and are well-suited for GPU execution, thereby contributing less than $1\%$ to the overall SCF iteration time.

\subsection{Multi-Resolution Quadrature}~\label{sec: 4.2}
The pseudo-core density ($\nTilde_c(\bx) = \sum_{\br}\sum_a\nTilde^a_c(\bx-\bR^a-\bL_{\br})$) contributes to the evaluation of the integrals appearing in Eqs. \eqref{eq: double counting}, ~\eqref{eq: ionic forces PAW} and \eqref{eq: FE discretized Poisson problem} and requires special care. We note that the support of the pseudo core density is larger than the augmentation radius  and requires a high quadrature rule to ensure sufficient sampling of the data; this is in contrast to other terms like $\bTilde^a(\bx)$ and $V^a_{\text{z}}(\bx)$ which have compact support less than augmentation radius.
Consequently, an accurate numerical integration of terms involving $\nTilde_c(\bx)$ require a sufficiently high-order quadrature rule or refined mesh, which would significantly increase the computational cost.
To ensure efficient and accurate evaluation of the integrals, we introduce a multi-resolution quadrature rule for integrating terms involving $\nTilde_c(\bx)$.
Since $\nTilde^a_c(\bx-\bR^a)$ has major contribution from within $\Omega_a$, integrals of the form  $\int_{\Omega_p}f(\bx)\nTilde^a_c(\bx)d\bx$ are evaluated as:
\begin{align}
    \sum_{\br}\int_{\Omega_p}f(\bx)\nTilde^a_c(\bx-\bR^a-\bL_{\br})d\bx &\rightarrow \biggl[\sum_{\br}\sum_t\sum_{e,q} w_q f(\bx^e_q) \nTilde^a_c(\bx^e_q-\bR^a-\bL_{\br})\det\{ \bJ_{(e,t)} \}\biggr\rvert_{\widehat{\bx}_{q}} \nonumber\\
    &  - \sum_t\sum_{e\in e^a} \sum_q w_q f(\bx^e_q)\nTilde^a_c(\bx^e_q-\bR^a)\det\{ \bJ_{(e,t)} \}\biggr\rvert_{\widehat{\bx}_{q}} \nonumber \\
    &+ \sum_t\sum_{e\in e^a}\sum_{Q} w_Q f(\bx^e_{Q})\nTilde^a_c(\bx^e_{Q}-\bR^a) \det \{\bJ_{(e,t)} \}\biggr\rvert_{\widehat{\bx}_{Q}}\color{black}\biggr]
    \label{eqn:MultiResolution quad rule}
\end{align}
\begin{figure}[H]
    \centering
    \includegraphics[scale=1.2]{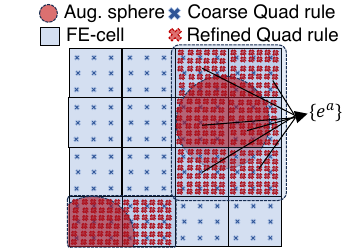}
    \caption{Multi-resolution quadrature}
    \label{fig:AdaptiveQuadrature}
\end{figure}
where $e^a$ refers to cells with support of the augmentation sphere $\Omega_a$, quadrature index '$q$' refers to a coarser quadrature rule which is usually used to evaluate integrals involving pseudo density $\nTilde(\bx)$ with a total of $n_{q_2}$ quadrature points. Further, the quadrature index '$Q$' refers to a refined quadrature rule that is usually the same as the one used to integrate terms involving compensation charge $\bTilde(\bx)$, which usually has around $21$ to $27$ total quadrature points in each direction.  As illustrated in Figure~\ref{fig:AdaptiveQuadrature}, cells which are part of the $\Omega_a$ support (denoted as $e^a$) are sampled with a refined quadrature rule '$Q$' while the other cells are sampled using a coarser rule.  
To this end, as demonstrated in Eqn.~\eqref{eqn:MultiResolution quad rule} for the cells which belong to $e^a$, we remove the contribution from the coarser quadrature rule (in blue) and add the contribution from the refined quadrature rule (in red). This enables accurate integration while avoiding unnecessary mesh refinement elsewhere, which is critical when using the PAW method with coarser FE meshes. 

\subsection{Efficient FE-Discretized PAW Operator Action on Trial Vectors}
\begin{figure}
    \centering
    \includegraphics[width=1.0\linewidth]{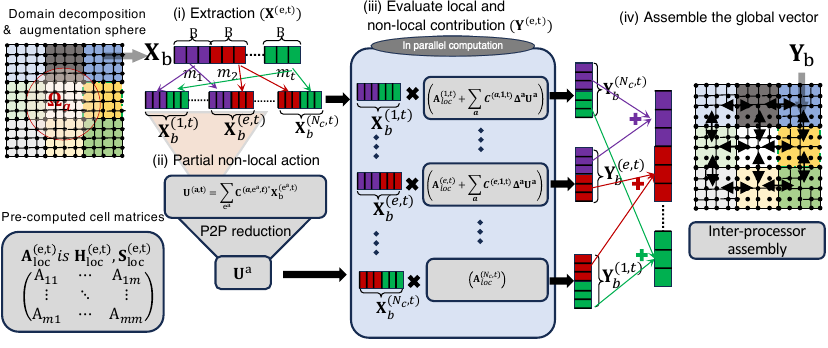}
    \caption{\textbf{Efficient computation of $\bA\bX$}: The computation of $\bY = \bA\bX$ involves 4 steps: (i) Extraction, (ii) Partial non-local operator action, (iii) Evaluation of local and non-local action and finally (iv) assembly of the global output vector. In the domain decomposition layout, each color is used to depict a unique MPI task (`$t$') and the degree of freedom is depicted with the dot symbol. Each task `$t$' is associated with `$N_c$' FE cells and `$m_t$' degrees of freedom. The vectors $\bX$ and $\bY$ are of dimension $m_t\times B$, where `$B$' is the multivector block size. }
    \label{fig: computing AX}
\end{figure}
In the proposed R-ChFSI approach in Algorithm~\ref{alg:ChFSI} discussed in the previous section, the efficient computation of matrix-multivector products of the form $\bA\bX$  is essential for enabling fast, accurate and scalable calculations on parallel computing architectures. We first consider the case of evaluating the matrix block-vector products of the form $\bA\bX$, where the matrix $\bA$ can be one of the discretised Hamiltonian ($\bH$), the PAW overlap ($\bS$) or its inverse. As shown in Figure~\ref{fig: computing AX}, the operator $\bA$ has two contributions $\bA_{\text{loc}}$ and $\bA_{\text{nloc}}$ . We note here that the matrix $\bA_{\text{loc}}$  is a large sparse matrix  of dimension $M \times M$ with sparsity dictated by the locality of the FE-basis functions while $\bX$ is a large dense matrix of dimension $M \times N$  with $N \approx \mathcal{O}(10) - \mathcal{O}(10^4)$ and $M \approx \mathcal{O}(10^4) - \mathcal{O}(10^8)$.  A naive strategy often adopted in the finite-element literature to compute $\bA\bX$ is to construct the FE discretised global sparse matrix $\bA$ using sparse-matrix storage data-structures and then compute the sparse matrix-dense matrix product $\bA\bX$~\cite{hughes2012finite,Kirby}. This strategy is inefficient on modern parallel computing systems involving multi-GPUs and multi-node architectures owing to the expensive data movement costs involved in the underlying sparse-matrix algorithms. To this end, we employ a cell-matrix approach~\cite{dftfe0.6,dftfe1.0,Kronbichler2012AApplication,JPDC} that avoids forming the global sparse matrix and instead uses dense–dense products at the FE-cell level to reduce memory requirements, enhance cache locality, and improve the arithmetic intensity of computation.
Figure~\ref{fig: computing AX} outlines the various steps involved in computing $\bA\bX$. The first step is the \textbf{extraction} of a block of wavefunctions to a cell-wise vector ($\bX^{(e,t)}_b$) which is of dimension $n_p^3\times B$, where $B\leq N$ is the multi-vector block size, $e$ denotes the cell index in the processor and $t$ denotes the corresponding task index. During the extraction step,  inter-processor communication is required to ensure that the contributions associated with degrees of freedom owned by neighboring processors (ghost nodes) are updated.

The second step involves computing the partial non-local contribution for atoms whose augmentation spheres overlap with the subdomain associated with a given MPI task , including all finite-element cells within the corresponding support. Figure~\ref{fig: computing AX} shows the cells and tasks that are part of the support of atom '$a$'.  The third step is to evaluate the local contribution and complete the remaining non-local action. Since this step is embarrassingly parallel, with computations associated with each finite-element cell being independent and requiring no synchronization, it is executed in parallel across GPU threads, with the output stored for each cell. Finally, a local assembly is performed to accumulate contributions from cells sharing the nodes, followed by an inter-processor accumulation to ensure the output field is continuous across MPI tasks, while setting the value of the discretized fields (entries of multivector X) at the ghost nodes to 0. Here, ghost nodes denote nodes that are shared across processors but are not owned by the current processor.
This is performed for all blocks $b \leq N_B$ of multivectors.
We describe the key mathematical aspects involved in this implementation strategy in Supporting Information (sec. S-3). 
The operation of $\bY = \bA\bX$ is performed blockwise, with each block comprising $B$-wavefunctions as elucidated before. This is done to reduce the peak memory requirement on GPUs and enables the design of compute-communication overlap strategies during the computation of $\bA\bX$.

Below, we describe the algorithmic innovations employed to build the Chebyshev-filtered subspace $\Psips_{\text{f}}$, leveraging the strategy for computing $\bA\bX$ discussed so far.

\subsection{Algorithmic Innovations in Subspace Construction}
As illustrated in Algorithm~\ref{alg: ChFSI}, the subspace construction scales as $\mathcal{O}(MN)$ in arithmetic complexity and is usually the primary computational bottleneck for systems up to 10,000 electrons. To this end, we discuss the various algorithmic innovations employed in this work to reduce the prefactor of $\mathcal{O}(MN)$ arithmetic complexity incurred in the construction of the Chebyshev-filtered space.  Recalling Eqns.~\eqref{eq: RChFSI residualfinal 0} and~\eqref{eq: RChFSI residual 1}, the recurrence relations to be evaluated in R-ChFSI are
\begin{align}
    \bR_{\theta+1} &= \frac{2\zeta_{\theta+1}}{e}\left(\bH\widehat{\bS}^{-1}\right)\bR_{\theta} -\frac{2\zeta_{\theta+1}c}{e}\bR_{\theta} - \zeta_{\theta}\zeta_{\theta+1}\bR_{\theta-1} + \frac{2\zeta_{\theta+1}}{e}\left(\bH\Psips_{\text{in}}-\bS\Psips_{\text{in}}\boldsymbol{\Lambda}_\text{in}\right)\boldsymbol{\Lambda}_{\theta} \label{eq: Recurrenace RChFSI dominant} \\
    \boldsymbol{\Lambda}_{\theta+1} &= \frac{2\zeta_{\theta+1}}{e}\boldsymbol{\Lambda}_{\text{in}}\boldsymbol{\Lambda}_{\theta} -\frac{2\zeta_{\theta+1}c}{e}\boldsymbol{\Lambda}_{\theta} - \zeta_{\theta}\zeta_{\theta+1}\boldsymbol{\Lambda}_{\theta-1}
\end{align}
Hence, the key kernel that has to be efficiently evaluated is of the form
\begin{equation}
    \bY = \alpha(\bH\widehat{\bS}^{-1})\bX + \beta \bX + \gamma \bY
    \label{eq: HXCheby}
\end{equation}
where $\alpha,\beta,\gamma$ are scalars that arise in the Chebyshev recurrence relation and are used to denote $\frac{2\zeta_{\theta+1}}{e}$, $\frac{2\zeta_{\theta+1}c}{e}$, $\zeta_{\theta}\zeta_{\theta+1}$ respectively, as defined in Algorithm~\ref{alg: CF: Residual Cheby}. Note that $\beta$ and $\gamma$ absorb the minus signs from the recurrence; the actual coefficients appearing in Eqns.~\eqref{eq: RChFSI residualfinal 0} and~\eqref{eq: RChFSI residual 1} are $-\beta$ and $-\gamma$. Referring to  Eqn.~\eqref{eqn:Sinvapprox} for the expression for $\widehat{\bS}^{-1}\bX$, the above expression is expanded as:
\begin{equation}
 \bY =   \alpha\bH \left[\bI - \sum_a{{\bM_{\text{D}}^{-1}}\bC^a{\boldsymbol{\Delta}^a _{\bS_{\text{inv}}}}\bC^{a\dagger}} \right]\bM_{\text{D}}^{-1}\bX + \beta \bX + \gamma \bY
 \label{eq: HXCheby_1}
\end{equation}
defining $\bT = \widehat{\bS}^{-1}\bX = \left[\bI - \sum_a{{\bM_{\text{D}}^{-1}}\bC^a{\boldsymbol{\Delta}^a _{\bS_{\text{inv}}}}\bC^{a\dagger}} \right]\bM_{\text{D}}^{-1}\bX$, Eqn.~\eqref{eq: HXCheby_1} is evaluated as a combination of $\bA\bX$ kernels, one to evaluate $\bT$ and another to evaluate $\bH\bT$.
To this end, the Supporting Information (sec. S-4) describes an efficient strategy for evaluating the recursion in Eqn.~\eqref{eq: HXCheby} by utilizing efficient $\bA\bX$ kernels discussed previously to evaluate $\bH\bX$ and $\widehat{\bS}^{-1}\bX$. Algorithm S-2 presented in Supporting Information (sec. S-4) takes  $\bX,\bY$ as inputs, while using  $\bT$ as a temporary variable to compute  $\bY$ as the output. Furthermore, this section S-4 prescribes in detail the various computation and communication steps involved in evaluating Eqn.~\eqref{eq: HXCheby}.

\begin{figure}[H]
    \centering
    \includegraphics[scale=1.0]{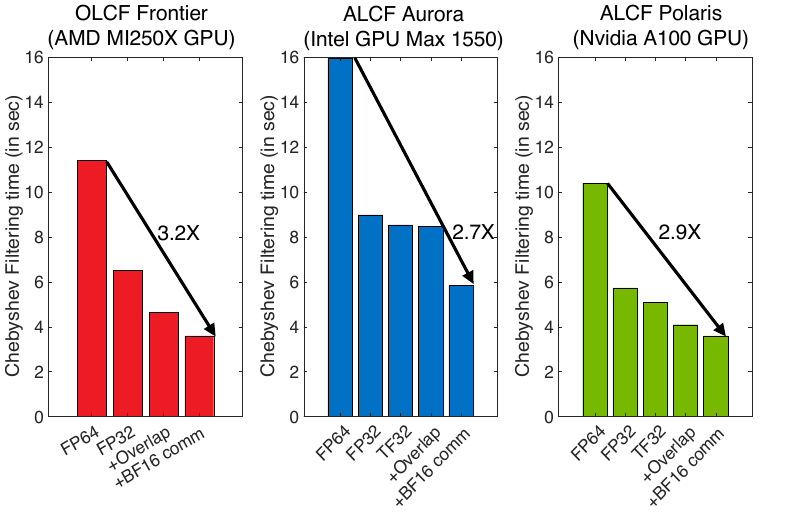}
    \caption{\textbf{Walk-through of algorithmic innovations for Chebyshev filtering}: comparison of the various algorithmic innovations on various supercomputers: OLCF Frontier, which uses AMD MI250X GPUs, ALCF Aurora, which uses Intel GPUs, and ALCF Polaris, which uses NVIDIA A100 GPUs. Benchmark system considered is Te deposited on WS\textsubscript{2} slab comprising 10,000 electrons and 6 million DoFs run on 120 GPUs where 70\% parallel-scaling efficiency is obtained.}
    \label{fig:WalkThrough}
\end{figure}

We now discuss the various algorithmic strategies employed to reduce the cost of constructing the Chebyshev filtered subspace. We will use Algorithm S-2 presented in the Supporting Information (sec. S-4) as the framework for discussing and benchmarking these strategies, which involve reduced-precision computation, overlapping compute and communication steps, and reduced-precision communication strategies.
 The performance gains of these strategies are demonstrated on a benchmark system consisting of a chain of eight Te atoms adsorbed on a WS\textsubscript{2} slab, comprising approximately 10,000 electrons and 6 million finite-element basis functions. The simulation is executed on 120 GPUs,  ($\approx$70\% parallel scaling efficiency) on leadership-class supercomputers including OLCF Frontier (AMD MI250X GPUs), ALCF Aurora (Intel GPUs), and ALCF Polaris (NVIDIA A100 GPUs). 
 Figure~\ref{fig:WalkThrough} illustrates the performance gains attained from each strategy elaborated in this subsection. 
\paragraph{Mixed precision computation:} Modern computing hardware is increasingly optimized for machine-learning workloads, providing high-throughput support for reduced-precision formats such as \texttt{BF16}, \texttt{FP8}, and \texttt{FP32}. Motivated by this trend, we seek to exploit low-precision compute and data-movement in our work to enable efficient large-scale DFT calculations. To this end, we note that the \texttt{CF}-step in the R-ChFSI algorithm used in \pawfe~is formulated in terms of the eigenproblem residuals and hence is tolerant to approximations in matrix-vector products. In the spirit of our recent work on norm-conserving pseudopotentials~\cite{RChFSI}, we use \texttt{FP32} arithmetic to compute the filtered space without sacrificing numerical accuracy or robustness. 
In this regard, for the recurrence relation defined in Eqn.~\eqref{eq: Recurrenace RChFSI dominant}, the contribution of $\left[  \frac{2\zeta_{\theta+1}}{e}\left(\bH\widehat{\bS}^{-1}\right)\bR_{\theta} -\frac{2\zeta_{\theta+1}c}{e}\bR_{\theta} - \zeta_{\theta}\zeta_{\theta+1}\bR_{\theta-1}\right]$ is evaluated using \texttt{FP32} arithmetic, while the addition of $\left[ \frac{2\zeta_{\theta+1}}{e}\left(\bH\Psips_{\text{in}}-\bS\Psips_{\text{in}}\boldsymbol{\Lambda}_\text{in}\right)\boldsymbol{\Lambda}_{\theta}\right]$ is performed in full-precision and the result($\bR_{\theta+1}$) is stored in \texttt{FP32} arithmetic.
 
From Figure~\ref{fig:WalkThrough}, employing the mixed precision compute strategy discussed here
we observe close to $1.8\times$ reduction in computational time in the \texttt{CF}-step across all supercomputers considered in this study.

\begin{figure}[H]
    \centering
    \includegraphics[scale=0.4]{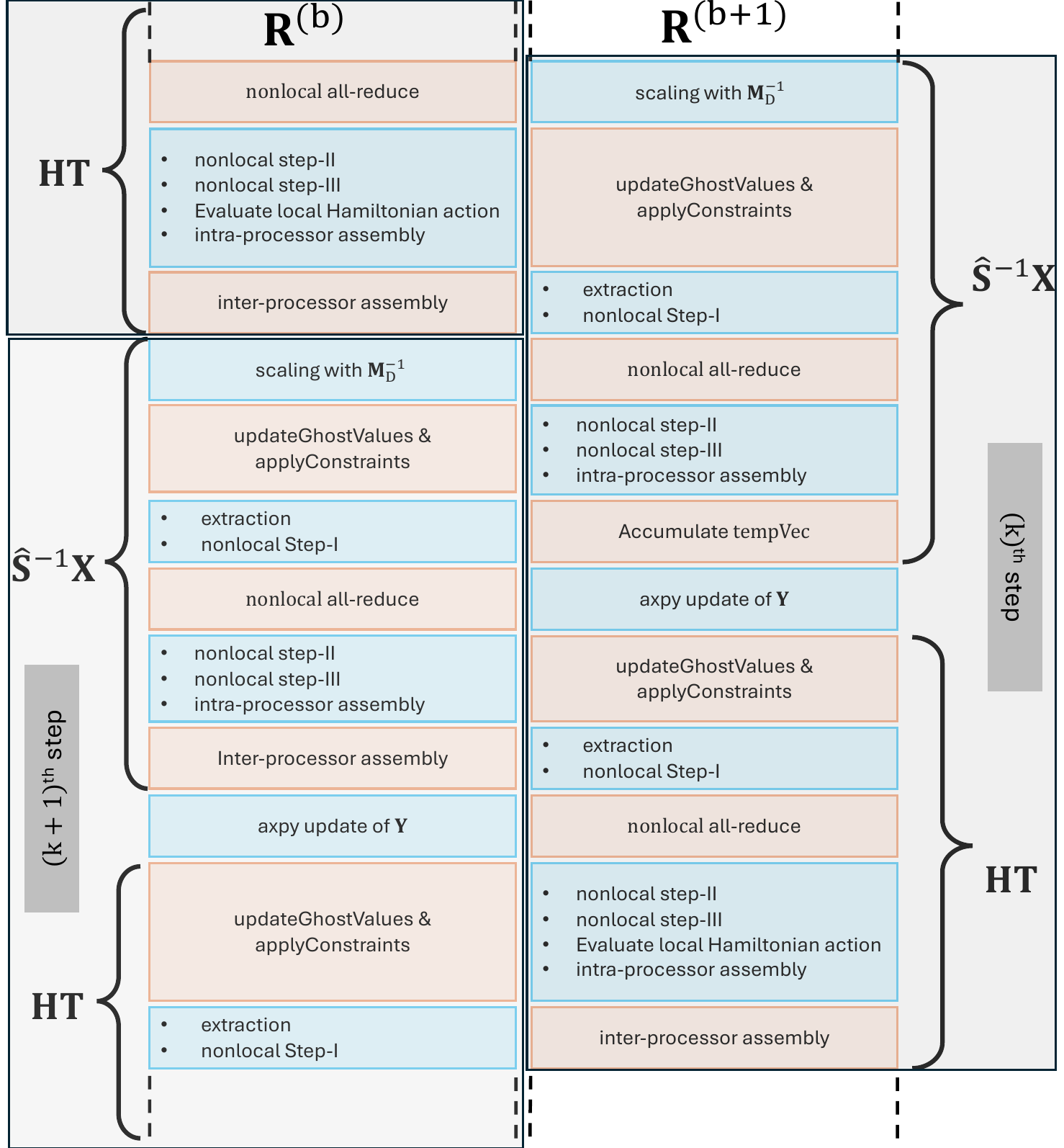}
    \caption{Schematic demonstrating the compute-communication overlap in the blocked R-ChFSI filtering step. The orange blocks involve inter-processor communication, while the blocks in blue involve computation. For each iteration, the action of $\widehat{\bS}^{-1}\bX$ and $\bH\bX$ for each block has to be completed. }
    \label{fig:BlockRChFSI}
\end{figure}

\begin{figure}[H]
    \centering
    \includegraphics[scale=1.0]{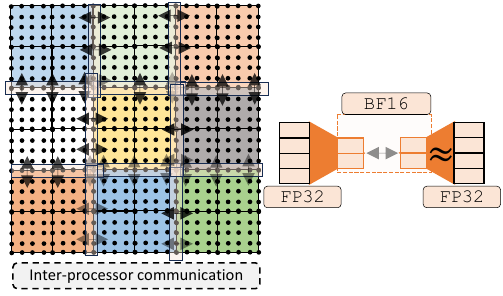}
    \caption{ \textbf{Reduced precision communication for nearest neighbour communication}: The figure illustrates the processor-level domain decomposition of the simulation domain, where each color represents a distinct processor. The arrows depict the inter-processor communication, which is performed in \texttt{BF16} format and cast to \texttt{FP32} for computation in the receiving processor.  }
    \label{fig:BF16Communication}
\end{figure}

 \paragraph{Overlap compute and communication in Chebyshev filtering:} One of the key advantages of the Chebyshev filtering approach is that the computation of the filtered space can be performed blockwise, without any coupling across blocks. Considering two blocks of trial vectors, each of size $B$, the successive updates in the recurrence relation can be evaluated as

\begin{align}
    &\bR^{(b)}_{\theta+1} = \frac{2\zeta_{\theta+1}}{e}\left(\bH\widehat{\bS}^{-1}\right)\bR^{(b)}_{\theta} -\frac{2\zeta_{\theta+1}c}{e}\bR^{(b)}_{\theta} - \zeta_{\theta}\zeta_{\theta+1}\bR^{(b)}_{\theta-1} + \frac{2\zeta_{\theta+1}}{e}\left(\bH\Psips^{(b)}_{\text{in}}-\bS\Psips^{(b)}_{\text{in}}\boldsymbol{\Lambda}^{(b)}_\text{in}\right)\boldsymbol{\Lambda}^{(b)}_{\theta}\nonumber \\
    &\boldsymbol{\Lambda}^{(b)}_{\theta+1} = \frac{2\zeta_{\theta+1}}{e}\boldsymbol{\Lambda}^{(b)}_{\text{in}}\boldsymbol{\Lambda}^{(b)}_{\theta} -\frac{2\zeta_{\theta+1}c}{e}\boldsymbol{\Lambda}^{(b)}_{\theta} - \zeta_{\theta}\zeta_{\theta+1}\boldsymbol{\Lambda}^{(b)}_{\theta-1} \;\text{and}\; \\
 &\bR^{(b+1)}_{\theta} = \frac{2\zeta_{\theta}}{e}\left(\bH\widehat{\bS}^{-1}\right)\bR^{(b+1)}_{\theta-1} -\frac{2\zeta_{\theta}c}{e}\bR^{(b+1)}_{\theta-1} - \zeta_{\theta-1}\zeta_{\theta}\bR^{(b+1)}_{\theta-2} + \frac{2\zeta_{\theta}}{e}\left(\bH\Psips^{(b+1)}_{\text{in}}-\bS\Psips^{(b+1)}_{\text{in}}\boldsymbol{\Lambda}^{(b+1)}_\text{in}\right)\boldsymbol{\Lambda}^{(b+1)}_{\theta-1}\nonumber \\
    &\boldsymbol{\Lambda}^{(b+1)}_{\theta} = \frac{2\zeta_{\theta}}{e}\boldsymbol{\Lambda}^{(b+1)}_{\text{in}}\boldsymbol{\Lambda}^{(b+1)}_{\theta-1} -\frac{2\zeta_{\theta}c}{e}\boldsymbol{\Lambda}^{(b+1)}_{\theta-1} - \zeta_{\theta-1}\zeta_{\theta}\boldsymbol{\Lambda}^{(b+1)}_{\theta-2}
\end{align}
 where '$\theta$' refers to the recurrence index denoting the $\theta^{th}$ step in building the filtered space using a Chebyshev polynomial of degree $m$ from $\theta=1$ to $m$. Furthermore, '$b$' refers to the block index of the trial block-vectors. As discussed in the previous section, the $\bA\bX$ kernel requires nearest-neighbor communication during the extraction and assembly steps.
In addition,  the non-local action in the $\bA\bX$ kernel requires communication between processors that own the compact support of the augmentation sphere. These communication operations are performed using non-blocking point-to-point MPI communication calls (\texttt{MPI\_Irecv} and \texttt{MPI\_Isend}) while ensuring that no bottlenecks arise along the communication pipeline.  
Modern computing hardware is increasingly compute-dense but often suffers from scaling limitations arising from stalls while waiting for data during inter-processor communication. To mitigate this imbalance, we employ a compute–communication orchestration strategy as illustrated in Figure~\ref{fig:BlockRChFSI}. In Figure~\ref{fig:BlockRChFSI},  each rectangular box (blue and orange rectangles) denotes the steps involved in algorithm S-2 and algorithm S-3 presented in Supporting Information (sec. S-4). The figure depicts two block-vectors ($\bR_b$ and $\bR_{b+1}$) for which the  Chebyshev-filtered space is constructed in parallel: while one block performs communication on the dedicated \texttt{communication}-stream (shown in orange block), the other simultaneously executes computation on the \texttt{compute}-stream (shown in blue). This overlap ensures that communication latency is effectively hidden, enabling improved scalability and better utilisation of available hardware resources. From the plots present in Figure~\ref{fig:WalkThrough}, we observe that the proposed overlap strategy results in 20\% 20\% gains in performance on AMD and NVIDIA based platforms over the mixed precision strategy discussed before and about $2.5\times$ over the baseline double precision implementation without overlapping compute and communication. We also find that, while both NVIDIA and AMD GPUs benefit from the overlap strategy, Intel GPUs show no noticeable improvement. We believe this is because the Intel GPU software stack is not yet mature enough to effectively exploit parallel \texttt{stream} execution, limiting the gains from the overlap strategy.
\paragraph{Reduced precision inter-processor communication:}
As observed in Figure~\ref{fig:BF16Communication}, the fraction of FE-nodal DoFs involved in inter-processor communication is quite small. To further reduce the volume of data exchanged across processors, we perform communication using the \texttt{BF16} data type. Furthermore, since communication is limited by bandwidth, especially for medium-scale to large-scale calculations, we expect a significant reduction in communication time, thereby ensuring better parallel scaling efficiency. Owing to the robustness of R-ChFSI procedure for constructing the filtered subspace, we observe that employing \texttt{BF16} precision for communication neither increases the number of iterations nor compromises robustness or accuracy. From the plots in Figure~\ref{fig:WalkThrough}, the synergy of all the innovations results in approximately a $3\times$ reduction in computational time for Chebyshev filtering construction across all available GPU architectures examined in this work, reducing the pre-factor that governs the quadratic scaling of filtered space construction.

 \subsection{Computational Strategies for the Rayleigh-Ritz Procedure}
The Rayleigh-Ritz (\texttt{RR}) procedure (steps 4-6 in Algorithm~\ref{alg:ChFSI}) starts becoming the computational bottleneck for large-scale systems comprising more than 10,000 electrons. Though the various steps in \texttt{RR} scale as $\mathcal{O}(N_e^3)$, from our benchmarks, the decreasing order of computational cost in \texttt{RR} is as follows: (i)\texttt{RR-GHEP}, (ii) \texttt{RR-SR}, (iii) \texttt{RR-Projection}. 

\begin{figure}[H]
    \centering
    \includegraphics[width=0.9\linewidth]{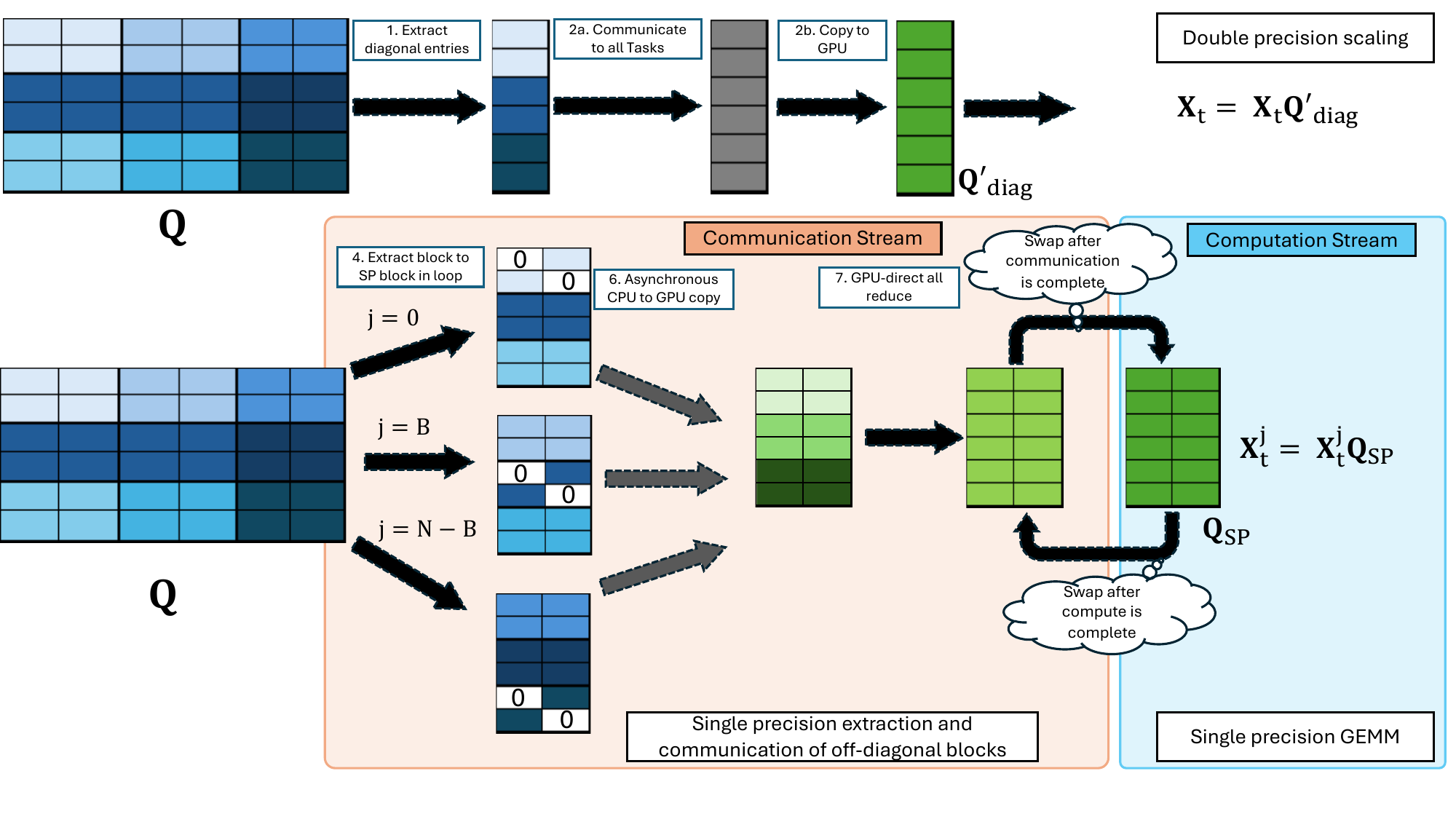}
    \caption{\textbf{Mixed precision subspace rotation}: Schematic of mixed-precision subspace rotation (\texttt{RR-SR}) strategy employed. The rotation matrix ($\bQ$) is distributed over various MPI tasks, and it is first communicated to all the MPI tasks using an \texttt{AllReduce} operation.
    The first row demonstrates the full precision multiplication of the process local dofs($\bX_t$) with the diagonal entries of the rotation matrix($\bQ$). The second row shows the off-diagonal multiplication of the local dofs with the rotation matrix, with diagonal entries set to zero ($Q_{ii}=0$). Here, the rotation matrix is extracted and communicated in single-precision and copied to the GPUs blockwise. There is an overlap between the communication of the subspace rotated block and the \texttt{GEMM} operation on GPUs to compute the rotation. The \texttt{communStream} is depicted in orange, and the \texttt{computeStream} is depicted in blue.  }
    \label{fig: mixed prec RR_SR}
\end{figure}

\paragraph{\texttt{RR-Projection}:} We outline the computational strategy employed to construct the projected Hamiltonian ($\bH^p$) and projected PAW overlap ($\bS^p$) matrices, both of which are computed using the operation $\bX^{\dagger}\bY$, where $\bY = \bH\bX$ or $\bY=\bS\bX$. To this end, $\bX$ is partitioned across MPI tasks $t$ according to the finite-element domain decomposition, yielding local blocks $\bX_t$. Similar in spirit to our previous works~\cite{pawfe,dftfe1.0}, the processor level $\bA^p_t = \bX_t^{\dagger}\bY_t$ is computed locally in each processor after which the global matrix $\bA^p$ is assembled by summing these contributions using an \texttt{MPI\_AllReduce} collective. Furthermore, to reduce the peak memory usage, only the lower triangle of $\bA^p_t$ is evaluated, and the computation is performed in blocks of size $N\times B_v$, where $B_v$ refers to the choice of wavefunction blocksize. Furthermore, the local computation of $\bA^p_t$ is performed on the \texttt{compute}-stream while the communication of $\bA^p_t$ to build $\bA^p$ is done on the \texttt{communication}-stream, thereby achieving compute–communication overlap. 
\paragraph{\texttt{RR-GHEP}:} Towards scalable and efficient evaluation of the eigenvalue problem $\bH^p\bQ=\bS^p\bQ\boldsymbol{\Lambda}$, the projected matrices $\bH^p,\bS^p$ are stored in a 2D-block-cyclic format. We employ the state-of-the-art \texttt{ELPA} library \cite{elpa1,elpa2} to solve the projected GHEP on GPUs or CPUs, with \texttt{OpenMP} parallelisation based on system size. The use of GPU parallel \texttt{GEMM} in \texttt{ELPA} enables efficient scaling on multi-node GPUs. After computing the eigenpairs, we perform the subspace rotation step to recover the eigenvectors in the full finite-element space.
\paragraph{\texttt{RR-SR}:}
 A mixed precision strategy, as shown in Figure~\ref{fig: mixed prec RR_SR}, is employed in \pawfe~without compromising on SCF convergence. \texttt{RR-SR} involves the computation of $\bX_t\bQ$, where the entries of $\bQ$ are distributed over various MPI-tasks. The key steps in \texttt{RR-SR} involve the extraction and copying of the entries of $\bQ$ and the subsequent \texttt{GEMM} operation to get the rotated space. As shown in Figure~\ref{fig: mixed prec RR_SR}, the two steps are overlapped over blocks of $\bQ$, with the extraction and copying occurring on a \texttt{communication}-stream (shown in orange) and the \texttt{GEMM} operation in the \texttt{compute}-stream. Additionally, these operations are performed in reduced precision (\texttt{FP32}) for the off-diagonal entries, reducing the overall cost. This is justified by the fact that $\bQ$ approaches the identity matrix as the SCF procedure converges, implying that the off-diagonal entries become progressively smaller. Consequently, evaluating these terms in reduced precision has a negligible effect on the rotated eigenspace. Furthermore, the \texttt{GEMM} operation is done in batches of DoFs to reduce the peak memory requirement. 

Furthermore, to validate that the various strategies and approximations employed in \pawfe~are robust and do not lead to any loss of accuracy, Table~\ref{Tab: accuracy of approximation} compares the error in converged ionic forces and ground-state energy on representative systems. We consider two representative benchmark systems:  (a)  Tellurium (Te) atoms deposited on WS\textsubscript{2} slab (semi-periodic) comprising  1208 atoms and 10,440 electrons, (b) Cu\textsubscript{309} nanoparticle (non-periodic) comprising  309 atoms and 5871 electrons. For both systems, the table reports the error with respect to the baseline calculations of full-precision R-ChFSI, without overlapping compute-communication and without mixed-precision strategies in either subspace construction or subspace rotation. Specifically, the table compares the accuracy of two reduced-precision configurations: (i) mixed-precision computation and communication (\texttt{FP32} compute and \texttt{BF16} communication) combined with compute–communication overlap in the Chebyshev filtering (\texttt{CF}) step, and (ii) the same strategy as in (i), with the additional use of mixed-precision computation and communication in the subspace rotation (\texttt{RR-SR}) step. From the table we observe that the error in the free energy is $\mathcal{O}(10^{-11})\frac{\text{Ha}}{\text{atom}}$ and ion-force is $\mathcal{O}(10^{-7})\frac{\text{Ha}}{\text{bohr}}$ which are orders of magnitude lower than the discretization error we seek for chemical accuracy.

\begin{table}
    \centering
    \begin{tabular}{|c|c|c|c|}
         \hline
         System & Algorithmic strategy & Energy error ($\frac{\text{Ha}}{\text{atom}}$)  & Force error ($\frac{\text{Ha}}{\text{bohr}}$)  \\
         \hline
      Cu\textsubscript{5-shell} Nanocluster   & mixed precision \texttt{CF} & $1.62\times 10^{-11}$ & $2.92\times 10^{-8}$  \\
      \cline{2-4}
       (309 Atoms)  & mixed precision \texttt{CF}\&\texttt{RR-SR} & $1.30\times 10^{-11}$ & $3.09\times 10^{-8}$  \\
         \hline
      Te deposited on WS\textsubscript{2}    & mixed precision \texttt{CF} & $7.71\times 10^{-12}$ &   $1.07\times 10^{-7}$\\
     \cline{2-4}
      (1208 atoms)   & mixed precision \texttt{CF}\&\texttt{RR-SR} & $2.47\times 10^{-11}$ & $4.9\times 10^{-8}$  \\
         \hline
    \end{tabular}
    \caption{Accuracy comparison of the various computational optimizations. Benchmark system: (a) Cu\textsubscript{5-shell} nanocluster comprising 309 atoms; (b) Te deposited on WS\textsubscript{2} comprising 1208 atoms. The energy error (in $\frac{\text{Ha}}{\text{atom}}$) and force error (in $\frac{\text{Ha}}{\text{bohr}}$) are computed relative to a baseline of R-ChFSI with full precision (in both \texttt{CF} and \texttt{RR-SR}) without overlapping compute and communication. In mixed precision \texttt{CF} only the filtered space construction uses \texttt{FP32} for compute and \texttt{BF16} for communication while overlapping compute and communication across wavefunction blocks.  While in mixed precision \texttt{CF}\&\texttt{RR-SR}, the \texttt{RR-SR} step is also performed with mixed precision compute-communication.   }
\label{Tab: accuracy of approximation}
\end{table}
\cb 
\subsection{Memory Footprint Reduction}

To enable nanoscale DFT simulations using \pawfe~on single-node and cloud-based computing resources, the peak memory footprint must be reduced. The most memory-intensive data structures in \pawfe~are the full wavefunction array, which scales as $N_{\mathrm{dof}} \times N$, the block wavefunction arrays used during Chebyshev filtering and the Rayleigh-Ritz procedure, which scales as $N_{\mathrm{dof}} \times B$ where $B$ is the multivector block size, and the cell-wavefunction array employed in the evaluation of $\bH\bX$, $\bS\bX$, and the electron density, which scales as $N_c \times n_p^3 \times B$ where $N_c$ denotes the total number of cells in the task.

One approach to reducing the peak memory footprint is to decrease the wavefunction block size ($B$), as discussed in the previous section. A complementary approach is to introduce a cell batch size, $N_c^B$, whereby finite elements are processed in batches rather than simultaneously. This reduces the memory requirement of the cell-wavefunction array from $N_c \times n_p^3 \times B$ to $N_c^B \times n_p^3 \times B$, while leaving the underlying algorithm unchanged. Throughout this work, we set $N_c^B = 50$, which was found empirically to provide a good balance between memory footprint and arithmetic intensity.

\subsection{Software Implementation}
To ensure \pawfe~is able to leverage the leadership class supercomputers 
 , the GPU implementation adopts a layered software architecture that separates hardware-independent algorithms from vendor-specific implementations. The majority of the codebase, including the self-consistent field iteration, eigensolver workflow, finite-element assembly, and parallel data structures, is written in a hardware-agnostic manner.

Platform-specific functionality is encapsulated within a thin abstraction layer responsible for GPU memory management, device synchronization, and performance-critical linear algebra kernels. This layer provides backend implementations for \texttt{CUDA}, \texttt{HIP}, and \texttt{oneAPI-SYCL}, exposing a common interface to the rest of the code. Consequently, higher-level algorithms remain unchanged irrespective of the target GPU architecture, while only the backend implementation requires adaptation to the underlying programming model.

This design minimizes code duplication, simplifies maintenance, and enables PAW-FE to support NVIDIA, AMD, and Intel GPU platforms through backend-specific implementations of a relatively small number of core routines.
\cn

 The following section demonstrates the impact of the computational methods and algorithmic innovations discussed thus far. Specifically, it presents the computational efficiency and parallel scalability of our work against state-of-the-art plane wave codes and against DFT-FE based on norm conserving pseudopotentials.

%% file: results.tex
In this section, we systematically evaluate the accuracy, performance, and scalability of the \pawfe~method on multinode GPU architectures. We begin by benchmarking the accuracy of spin-polarised implementation in \pawfe~against plane-wave methods in predicting ground-state energies and ionic forces, thereby establishing its reliability for first-principles calculations. We then assess the performance of \pawfe~on modern GPU architectures by quantifying the computational gains relative to CPU-based executions. Given the increasing dominance of GPU-accelerated platforms in both leadership-class supercomputing and on-demand cloud environments, such an assessment is critical to understanding potential reductions in time-to-solution, computational cost, and energy consumption in \textit{ab initio} simulations of systems with more than 3000 electrons. Subsequently, 
we compare the computational cost and minimum wall time of \pawfe~with Quantum ESPRESSO (\qe)\cite{qe}, a widely used plane-wave DFT code, across non-periodic and semi-periodic systems. In addition, we benchmark the computational cost and parallel scaling efficiency of \pawfe~against \DFTFE~, which employs optimized norm-conserving Vanderbilt (ONCV) pseudopotentials~\cite{ONCV, dftfe1.0}, using a Pt nanoparticle supported on a SrTiO\textsubscript{3} slab. This system serves as a representative model for metal–oxide interfaces of broad relevance in heterogeneous catalysis, where strong metal–support interactions play a critical role in catalytic activity and stability~\cite{PtOnSrTiO3}.
Finally, to demonstrate the capability of \pawfe~for large length-scale simulations, we assess the computational cost per SCF iteration for a large-scale twisted bilayer WTe\textsubscript{2} system comprising $\sim$ 100,000 electrons. 
\subsection{Accuracy Benchmarking}
In this section, we validate the accuracy of the spin-polarized implementation of \pawfe~against \abinit~\texttt{v10.6.3}~\cite{abinit2025}, a widely used plane-wave based density functional theory code. In \abinit, we employ \texttt{usexcnhat}=0 to enforce the Bl\"ochl formulation of the exchange-correlation (XC) contribution and set \texttt{pawxcdev}=0 to avoid approximations in the XC correction term~\cite{Abinit2010}. To this end, we consider four non-periodic and four periodic benchmark systems spanning open-shell molecules and magnetic crystalline solids, thereby validating the spin-polarized PAW implementation for both isolated and periodic systems. The non-periodic systems comprise O\textsubscript{2}, NO\textsubscript{2}, CN, and OH molecules, while the periodic systems comprise the primitive unit cells of SiO\textsubscript{2}, Co, Fe, and Cr. A more comprehensive benchmarking study involving spin-unpolarized calculations has been presented in our previous work~\cite{pawfe}.

The PAW datasets employed in this work are obtained from the \texttt{JTH v2.0} repository~\cite{JTH_PBE_v2_0}. All calculations employ the generalized gradient approximation (GGA)~\cite{GGA} with the PBE exchange--correlation functional~\cite{PBE}. For \pawfe, the finite-element discretization parameters are chosen as $p=7$, $p_{\mathrm{el}}=9$, and mesh size $h=1.0$ bohr. In \abinit, we employ a wavefunction cutoff energy of $E^{\mathrm{wfc}}_{\mathrm{cut}}=60$ Ha and a charge-density cutoff of $E^{\mathrm{rho}}_{\mathrm{cut}}=300$ Ha.

\begin{figure}[H]
    \centering
    \includegraphics[scale=1.1]{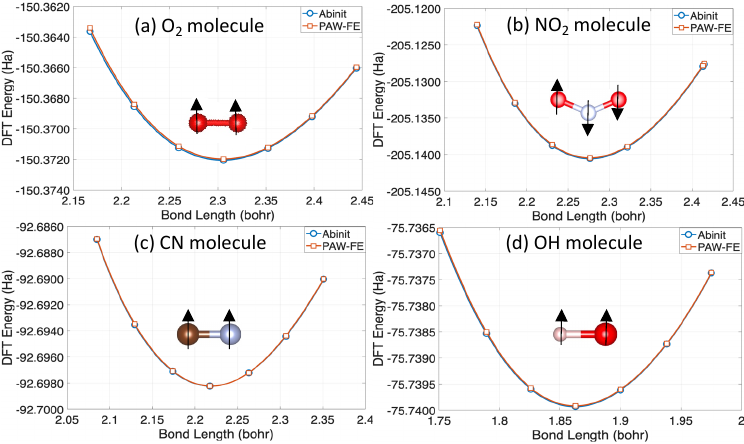}
    \caption{\textbf{Accuracy benchmarking}: comparison of energy against bond-length. Systems considered are: (a) O\textsubscript{2} molecule (b) NO\textsubscript{2} molecule (c) CN molecule (d) OH molecule. The DFT Energy in Ha is plotted on the Y-axis and the bond-length on the X-axis. The plot inset shows the structure considered and the initial spin configuration considered.}
    \label{fig:accuracyBM_GPU_NP}
\end{figure}

For the molecular systems, Figure~\ref{fig:accuracyBM_GPU_NP} compares the variation of the DFT total energy as a function of bond length. The calculations are performed in an orthorhombic simulation cell of dimensions $30~\mathrm{bohr}\times28~\mathrm{bohr}\times32~\mathrm{bohr}$. The equilibrium geometries are first obtained through ionic relaxation using \pawfe, yielding optimized bond lengths of $2.305$ bohr for the spin-triplet ground state of O\textsubscript{2}, $2.218$ bohr for CN, and $1.863$ bohr for OH. For NO\textsubscript{2}, the optimized geometry corresponds to a bond length of $2.276$ bohr and a bond angle of $134^\circ$. For each molecule, the bond length is varied from $94\%$ to $106\%$ of its equilibrium value, resulting in seven configurations used to construct the energy curves. The insets in Figure~\ref{fig:accuracyBM_GPU_NP} illustrate the corresponding molecular structures together with the initial magnetic configurations. O\textsubscript{2} is initialized in a spin-triplet state and converges to a net magnetization of $2.0~\mu_{\mathrm{B}}$ for all bond lengths considered. For NO\textsubscript{2}, the initial magnetic configuration corresponds to antiparallel alignment of the two oxygen atoms, yielding a converged net magnetization of $1.0~\mu_{\mathrm{B}}$. For the CN and OH molecules, a parallel spin configuration is employed as the initial guess, and both systems converge to a net magnetization of $1.0~\mu_{\mathrm{B}}$ in both \pawfe~and~\abinit. Furthermore, as shown in Figure~\ref{fig:accuracyBM_GPU_NP} and detailed in the Supporting Information (Section S-5), \pawfe~reproduces the \abinit~energy curves with excellent agreement over the entire range of bond lengths considered, with total-energy differences of the order of $10^{-5}~\mathrm{Ha/atom}$.

Finally, we validate the ionic forces computed by \pawfe~against \abinit. For the O\textsubscript{2} molecule at a bond length of $2.44$ bohr, the force error is $3.4\times10^{-5}~\mathrm{Ha/bohr}$; for NO\textsubscript{2} at a bond length of $2.62$ bohr, the error is $3.5\times10^{-5}~\mathrm{Ha/bohr}$; for CN at a bond length of $2.35$ bohr, the error is $4.01\times10^{-6}~\mathrm{Ha/bohr}$; and for OH at a bond length of $1.975$ bohr, the error is $4.63\times10^{-6}~\mathrm{Ha/bohr}$. These comparisons are performed for configurations in which the PAW augmentation spheres do not overlap, consistent with the assumptions underlying the present implementation.

\begin{figure}[H]
    \centering
    \includegraphics[scale=1.1]{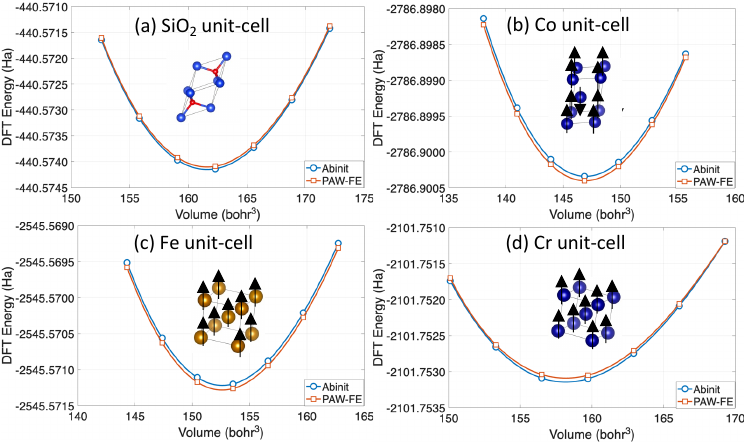}
    \caption{\textbf{Accuracy benchmarking}: comparison of energy against unit-cell volume. Systems considered are: (a) SiO\textsubscript{2} unit-cell (b) Co unit-cell (c) Fe unit-cell (d) Cr unit-cell. The DFT Energy in Ha is plotted on the Y-axis and the cell volume on the X-axis. The plot inset shows the structure considered and the initial spin configuration considered.}
    \label{fig:accuracyBM_GPU_Per}
\end{figure}

We next consider periodic benchmark systems. The reference structures of Cr, Fe, and Co are obtained from the $\Delta$-test repository~\cite{deltaTest}, with reference volumes of $159.70$~bohr\textsuperscript{3}, $153.52$~bohr\textsuperscript{3}, and $146.85$~bohr\textsuperscript{3}, respectively. For SiO\textsubscript{2}, we employ the primitive unit cell of the ideal cubic $\beta$-cristobalite phase with a reference volume of $162.32$~bohr\textsuperscript{3}. The unit-cell volume is varied from $94\%$ to $106\%$ of the corresponding reference value. Brillouin-zone sampling is performed using Monkhorst--Pack meshes of $12\times12\times12$ for Cr and SiO\textsubscript{2}, $15\times15\times15$ for Fe, and $15\times15\times8$ for Co. These $\mathbf{k}$-point grids are chosen such that they satisfy the $\Delta$-test sampling criterion of approximately $6000$ $\mathbf{k}$-points per reciprocal atom~\cite{deltaTest}. As shown in the insets of Figure~\ref{fig:accuracyBM_GPU_Per}, SiO\textsubscript{2} is initialized in a non-magnetic state, Fe and Co are initialized in ferromagnetic configurations, while Cr is initialized in an antiferromagnetic configuration. Figure~\ref{fig:accuracyBM_GPU_Per}, together with the results presented in the Supporting Information, demonstrates excellent agreement between \pawfe~and~\abinit~over the entire range of unit-cell volumes considered, with energy differences comparable to those observed for the molecular systems.

The excellent agreement in total energies, magnetic moments, and ionic forces demonstrates the accuracy of the spin-polarized PAW implementation in \pawfe. Having established the numerical accuracy of the method for both molecular and periodic systems, we next assess its computational performance on modern GPU-based supercomputing architectures.

\subsection{Performance Benchmarking}
In this sub-section, we demonstrate the performance gains achieved through the algorithmic innovations discussed in the previous sections. The performance comparisons are carried out on leadership-class supercomputing resources. Specifically, we consider (i) ALCF Aurora, where each node consists of six Intel GPUs with a total of 
768 GB of GPU memory, Intel Xeon processors, 1024 GB of system memory, and 
128 GB of high-bandwidth memory; (ii) OLCF Frontier, where each compute node 
comprises four AMD MI250X GPUs with a total of 512 GB of GPU memory, AMD EPYC 
7763 CPUs, and 512 GB of system memory; and (iii) ALCF Polaris, where each 
compute node comprises four NVIDIA A100 GPUs with 40 GB of memory each and an 
AMD EPYC Milan 7543P 32-core CPU with 512 GB of system memory.  These architectures represent state-of-the-art heterogeneous computing platforms with distinct GPU design and memory hierarchies. The PAW datasets are obtained from the recently released \texttt{JTHv2.0}\cite{JTH_PBE_v2_0} repository. All benchmarking studies presented here are performed using the generalized gradient approximation (GGA)~\cite{GGA} for the exchange–correlation functional, specifically employing the PBE formulation~\cite{PBE}. The finite element discretization parameters are such that the energies have error of $2\times10^{-4}\frac{\text{Ha}}{\text{atom}}$ in energy and $2\times10^{-4}\frac{\text{Ha}}{\text{Bohr}}$ in ionic forces. As mentioned in Sections~\ref{sec: 4.1} and~\ref{sec: 4.2}, the various quadrature rules are selected such that they introduce a higher order error in energy and ionic forces than the discretization error. Table~\ref{tab:fe_parameters_pawfe} lists the finite-element (FE) discretization parameters: FE interpolating polynomial degree for wavefunctions($p$), FE interpolating polynomial degree for electrostatic potential ($p_{\text{el}}$) and minimum mesh size ($h$). 
\begin{table}[h]
\centering
\begin{tabular}{|l|c|c|c|c|}
\hline
\textbf{System} 
& \textbf{Pt} & \textbf{Te deposited on} & \textbf{Pt nanocluster on} & \textbf{bi-layer } \\
& \textbf{nanocluster} 
& \textbf{WS\textsubscript{2} slab} 
& \textbf{SrTiO\textsubscript{3} slab} 
& \textbf{WTe\textsubscript{2} slab} \\
\hline
$p,p_{\text{el}},h$ & $6,6,2.0$  & $5,7,1.60$ & $7,8,1.90$  & $5,7,1.60$  \\
\hline
\end{tabular}
\caption{Finite-element discretization parameters used in \pawfe~ calculations for the various benchmark systems considered.}
\label{tab:fe_parameters_pawfe}
\end{table}

\subsubsection{CPU-GPU Speedup}
To quantify the performance gains of \pawfe~on GPU architectures relative to CPU-based executions, we consider two representative benchmark systems: (i) a Pt\textsubscript{561} nanocluster comprising 5,610 electrons and (ii) a Pt\textsubscript{1415} nanocluster comprising 14,150 electrons.

\cb Table~\ref{tab: CPU-GPU speedups} compares the wall time per self-consistent field (SCF) iteration for CPU and GPU executions on OLCF Frontier, ALCF Aurora, and ALCF Polaris. The baseline GPU implementation employs only \texttt{FP64} arithmetic and does not utilize the mixed-precision strategies or algorithmic optimizations introduced in the previous section. From this Table~\ref{tab: CPU-GPU speedups}, we observe that the baseline GPU implementation achieves CPU-GPU speedups of approximately $10\times$ on OLCF Frontier, $6\times$ on ALCF Aurora, and $8.8\times$ on ALCF Polaris relative to CPU-only executions (see column 6 of Table~\ref{tab: CPU-GPU speedups}). These improvements stem from both the computational characteristics of the PAW generalized Hermitian eigenproblem and the GPU-aware implementation strategies developed in this work. The dominant computational kernels in the Residual-based Chebyshev Filtered Subspace Iteration (R-ChFSI) algorithm (Algorithm~\ref{alg: ChFSI}) are (i) the Chebyshev-filtered subspace construction  (step 3) and (ii) the Rayleigh-Ritz procedure , comprising the projection (step 4), diagonalisation (step 5) and subspace rotation (step 6). These operations are dominated by dense matrix-matrix multiplications and other Level-3 \texttt{BLAS} kernels, which exhibit high arithmetic intensity and regular memory-access patterns, enabling efficient utilization of modern GPU architectures. Incorporating the efficient computational algorithms and GPU-centric optimizations proposed in this work further increases CPU-GPU speedups to approximately $18\times$, $9\times$, and $13\times$, respectively. This additional acceleration obtained over the baseline GPU implementation arises from the mixed-precision compute and communication strategies together with the computation-communication overlap introduced in this work. These optimizations reduce memory traffic and communication overhead while maintaining double-precision accuracy through the residual-correction mechanism of R-ChFSI, thereby improving overall device utilization and reducing time to solution.

The variation in speedups across Frontier, Aurora, and Polaris is primarily attributed to differences in the underlying CPU and GPU architectures, including the available memory bandwidth, computational throughput, interconnect characteristics, and the ratio of number of CPUs to GPUs on each system. Overall, these results demonstrate that combining hardware-aware numerical algorithms with architecture-specific implementation strategies is essential for achieving high-performance, scalable first-principles calculations on modern GPU-based supercomputing platforms.  \cn

\begin{table}[!h]
\centering
\footnotesize{
\begin{tabular}{|l|c|c|c|c|c|c|c|}
\hline
\multirow{2}{*}{Supercomputer} & \multirow{2}{*}{System} & \multirow{2}{*}{Nodes used} & CPU & GPU & Speed-up & GPU-optim & Speed-up\\
& & &Time(s) & Time(s) & \tiny {CPU/GPU} & Time(s) & \tiny{CPU/GPU-optim}\\
\hline
{OLCF Frontier } 
& Pt\textsubscript{561}  & 1 &373.35 & 36.23   & 10.3  & 21.21 & 17.8\\
\cline{2-8}
(AMD MI250X GPUs)& Pt\textsubscript{1415} & 3&887.84 & 82.63  & 10.7 & 46.52 & 19.1 \\
\hline
\hline
{ALCF Aurora } 
& Pt\textsubscript{561}  & 1&157.63 & 25.09 & 6.3 & 16.13& 9.8\\
\cline{2-8}
(Intel GPUs)& Pt\textsubscript{1415} & 2&478.92 &  87.11 & 5.5 & 58.34 & 8.2\\
\hline
\hline
\cb
{ALCF Polaris } 
& Pt\textsubscript{561}  & 2&221.27 & 25.42 & 8.7 & 15.92 & 13.9\\
\cline{2-8}
(NVIDIA A100 GPUs) & Pt\textsubscript{1415} & 7&470.64 & 53.67  & 8.8 & 34.76 & 13.5 \cn\\ 
\hline
\end{tabular}}
\caption{Comparison of average time per SCF iteration between CPU, GPU and the optimized GPU executions of \pawfe~on OLCF Frontier,  ALCF Aurora and \cb ALCF Polaris \cn for Pt\textsubscript{561} and Pt\textsubscript{1415} nanoclusters. The table also shows the minimum number of nodes required and the speedups relative to CPU execution. }
\label{tab: CPU-GPU speedups}
\end{table}

\subsubsection{Comparison with Plane-Wave Approaches}
We now proceed to compare the performance of \pawfe~against \qe, a plane-wave DFT code. To benchmark the computational efficiency of \pawfe~against \qe~, we choose benchmark systems comprising slabs (semi-periodic)  and nanoclusters (non-periodic) of varying sizes. In \pawfe, SCF convergence is determined using the energy residual defined by Eqn. S-19 in the Supporting Information. To ensure a consistent comparison, both \pawfe~and \qe~employ an energy convergence threshold of $2\times 10^{-8}$ Ha.

\begin{figure}[H]
    \centering
    \begin{subfigure}{0.45\textwidth}
        \centering
        \includegraphics[width=1.0\linewidth]{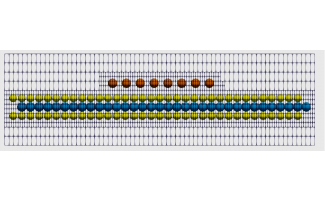}
        \caption{ \centering \textbf{Semi-periodic system} }
    \end{subfigure}%
    \begin{subfigure}{0.45\textwidth}
        \centering
        \includegraphics[width=1.0\linewidth]{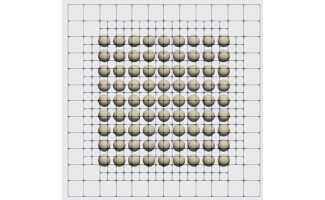}
        
        \caption{\centering \textbf{Non-periodic system} }
    \end{subfigure}
    \caption{\textbf{Finite-element meshes employed in \pawfe.} (a) Semi-periodic Te adsorbed on a WS\textsubscript{2} slab with a mesh size of 1.6 bohr in the vicinity of the atoms and a base mesh size of 3.2 bohr. (b) Non-periodic Pt nanocluster with a mesh size of 2.0 bohr around the atoms and a base mesh size of 4.0 bohr. Both meshes employ a two-level adaptive refinement strategy, with fine elements around the atoms and a coarser discretization in the vacuum region to reduce the overall number of degrees of freedom while maintaining accuracy.}
    \label{fig: FE mesh for PAW-FE QE comparison}
\end{figure}

The simulations reported here are performed on NVIDIA A100 GPUs on the ALCF Polaris supercomputer. The \qe~code (\texttt{v7.5}) employed in this study is compiled using the optimal compilers and libraries by the ALCF Polaris support team. The discretization parameters in both \pawfe~and \qe~ are chosen to ensure comparable accuracy in total energies and forces.  We refer to Figure~\ref{fig: FE mesh for PAW-FE QE comparison} for a representation of the finite-element mesh employed in \pawfe~for the performance benchmarking study in this section.
\paragraph{Semi-periodic systems.}
The semi-periodic benchmark system consists of WS\textsubscript{2} slabs of varying sizes with a chain of eight Te atoms adsorbed on the surface. The Te–slab separation is approximately $5.138$ bohr, while the W–S bond length is $4.53$ bohr. $\Gamma$-point sampling is employed for Brillouin zone integration. The slab is centered along the non-periodic direction with 10 bohr of vacuum on either side.
In \qe~, dipole correction is employed to account for artificial periodicity along the non-periodic direction~\cite{QEdipoleCorrection}, whereas in \pawfe~, zero-Neumann boundary conditions are imposed for the electrostatic potential~\cite{electricFieldJCTC}.
In this study we use plane-wave ($E^{\text{wfc}}_{\text{cut}},E_{\text{cut}}^{\text{rho}}$) and FE ($p,p_{\text{el}},h$) discretization parameters such that the energy error and force error relative to a refined calculation is $ \mathcal{O}(10^{-4})\frac{\text{Ha}}{\text{atom}}$ and $\mathcal{O}( 10^{-4})\frac{\text{Ha}}{\text{bohr}}$.
The finite-element discretization parameters are reported in Table~\ref{tab:fe_parameters_pawfe}, while  for \qe~, we use a kinetic energy cutoff of $E^{\text{wfc}}_{\text{cut}} = 25~\mathrm{Ha}$ for the wavefunctions and $E_{\text{cut}}^{\text{rho}} = 200~\mathrm{Ha}$ for the charge density.

Table~\ref{tab:WS2Te_performance} compares the number of basis functions, minimum number of nodes required and computational cost per SCF iteration ($\eta_\text{c}$) between \pawfe~and \qe. We observe that for systems larger than approximately 6000 electrons, \pawfe~demonstrates lower computational cost than \qe~, achieving over $3.6\times$ speedup for systems exceeding 10,000 electrons. Additionally, \pawfe~requires fewer computational resources (in terms of number of GPUs) to perform the same calculations. 

\begin{table}
\cb
\footnotesize{
\centering
\begin{tabular}{|c|c|c|c|c|c|}
\hline
System & No. of PWs/atom  & No. of DoFs/atom &$\eta_c$-\qe & $\eta_c$-\pawfe & Speedup  \\
(No. of electrons) & in \qe & in \pawfe &(Min. Nodes) & (Min. Nodes)  &\\
\hline
440 Atoms & \multirow{2}{*}{1,825} & \multirow{2}{*}{6,094} &0.008  & 0.011  & \multirow{2}{*}{0.75}  \\
(3792) & & &(2) & (2)& \\
\hline
776 Atoms & \multirow{2}{*}{1,840}&\multirow{2}{*}{5,789} & 0.043  & 0.026  & \multirow{2}{*}{1.67}  \\
(6704) & & &(4) & (3)& \\
\hline
1208 Atoms & \multirow{2}{*}{1,847} & \multirow{2}{*}{5,930}&0.210  & 0.058  & \multirow{2}{*}{3.63}  \\
(10440) & & &(12) & (6)&  \\
\hline
1736 Atoms &\multirow{2}{*}{1,851} & \multirow{2}{*}{6,033}& \multirow{2}{*}{--} & 0.124 & \multirow{2}{*}{--}  \\
(15024) & & & & (12) & \\
\hline
\end{tabular}
}
\caption{\cb \footnotesize{Comparison of the computational cost per self-consistent field (SCF) iteration ($\eta_c$) between \pawfe~and \qe~for semi-periodic WS\textsubscript{2} slabs with eight Te atoms adsorbed on the surface. The table reports the number of electrons, the basis size per atom (plane waves in \qe~and finite-element degrees of freedom in \pawfe), the minimum number of compute nodes required to satisfy the memory requirements (Min. Nodes), and the corresponding computational cost in node-hours ($\eta_c$-\qe~and $\eta_c$-\pawfe) on ALCF Polaris GPUs node. The number of KS orbitals solved in these calculations is approximately 60\% of number of electrons in the system.}
\cn}
\label{tab:WS2Te_performance}
\cn
\end{table}

\paragraph{Non-periodic systems.}
The non-periodic benchmark systems comprise cuboctahedral Pt nanoparticles of increasing size, with a nearest-neighbor distance of $5.4$ bohr. The system is centered within the simulation domain with 10 bohr of vacuum on either side. This vacuum size is chosen such that increasing it to 50 bohr results in differences of $\mathcal{O}(10^{-5})~\mathrm{Ha/atom}$ in energy and forces. In \pawfe~, zero-Dirichlet boundary conditions are imposed for the electrostatic potential, while periodic boundary conditions are employed in \qe. To accelerate SCF convergence, local Thomas-Fermi mixing with a mixing parameter of 0.02 is used in \qe, whereas \pawfe~employs adaptive Anderson mixing~\cite{Novak2023AdaptiveCalculations}. The finite-element discretization parameters are reported in Table~\ref{tab:fe_parameters_pawfe}, while \qe~calculations use $E^{\text{wfc}}_{\text{cut}} = 27.5~\mathrm{Ha}$ and $E_{\text{cut}}^{\text{rho}} = 170~\mathrm{Ha}$.

\begin{table}
\cb
\footnotesize{
\centering
\begin{tabular}{|c|c|c|c|c|c|}
\hline
System & No. of PWs/atom  & No. of DoFs/atom &$\eta_c$-\qe & $\eta_c$-\pawfe & Speedup  \\
(No. of electrons) & in \qe & in \pawfe &(Min. Nodes) & (Min. Nodes)  &\\
\hline
Pt\textsubscript{309} & \multirow{2}{*}{2,785} & \multirow{2}{*}{4,949} &0.006  & 0.003  & \multirow{2}{*}{1.89}  \\
(3090) & & &(2) & (1)& \\
\hline
Pt\textsubscript{561} & \multirow{2}{*}{2,396}&\multirow{2}{*}{4,678} & 0.029  & 0.01  & \multirow{2}{*}{3.01}  \\
(5610) & & &(5) & (2)& \\
\hline
Pt\textsubscript{923} & \multirow{2}{*}{2,051} & \multirow{2}{*}{4,366}&0.127  & 0.027  & \multirow{2}{*}{4.72}  \\
(9230) & & &(12) & (4)&  \\
\hline
Pt\textsubscript{1415} &\multirow{2}{*}{1,670} & \multirow{2}{*}{3,746}& \multirow{2}{*}{--} & 0.067 & \multirow{2}{*}{--}  \\
(14150) & & & & (7) & \\
\hline
\end{tabular}
}
\caption{\cb \footnotesize{Comparison of the computational cost per self-consistent field (SCF) iteration ($\eta_c$)  between \pawfe~and \qe~for non-periodic Pt nanocluster of various sizes. The table reports the number of electrons, the basis size per atom (plane waves in \qe~and finite-element degrees of freedom in \pawfe), the minimum number of compute nodes required to satisfy the memory requirements (Min. Nodes), and the corresponding computational cost in node-hours ($\eta_c$-\qe and $\eta_c$-\pawfe) on ALCF Polaris GPUs node. The number of KS orbitals solved in these calculations is approximately 60\% of number of electrons in the system.}
\cn}
\label{tab:PtNP_performance}

\end{table}

Table~\ref{tab:PtNP_performance} compares the computational cost per SCF iteration ($\eta_c$) between \pawfe~and \qe~. From this Table, we observe that \pawfe~consistently requires fewer computational resources than \qe~for the same system size. \pawfe~demonstrates lower computational cost for systems larger than 3000 electrons, with speedups approaching $4.7\times$ for systems close to 10,000 electrons. The Pt\textsubscript{1415} system was not feasible within the available memory and node limits using \qe~ on ALCF Polaris.  The improved computational efficiency of \pawfe~for non-periodic systems is primarily attributed to the spatial adaptivity of the finite-element basis, which allows coarsening in vacuum regions. In contrast, the plane-wave basis enforces uniform resolution across the entire simulation domain. 

Table~\ref{tab: Min Wall Time} reports the minimum wall time per SCF iteration ($\tau_c$) in seconds on ALCF Polaris. The wall time mentioned is measured at a parallel scaling efficiency of $25\%-30\%$. For both semi-periodic and non-periodic systems, \pawfe~achieves lower minimum wall time with about $3\times$ for system sizes around 3000 electrons and beyond $7\times$ for systems larger than 10,000 electrons. This is attributed to the locality of the finite-element basis, which enables nearest-neighbor communication, in contrast to the global all-to-all communication required in plane-wave methods, which scales poorly with increasing MPI concurrency.

\begin{table}[H]
\centering
\begin{minipage}{0.48\linewidth}
\centering
\subcaption{\textbf{Case Study:} Te deposited on WS\textsubscript{2} slab}
\footnotesize{\begin{tabular}{|c|c|c|c|}
\hline
System  & Electrons & $\tau$-\qe   & $\tau$-\pawfe   \\
 -& - & (\# nodes) & (\# nodes) \\
\hline
\multirow{2}{*}{440 Atoms} & \multirow{2}{*}{3792}  & 8.3 & 2.7  \\
& & (16) & (64) \\
\hline
\multirow{2}{*}{776 Atoms} & \multirow{2}{*}{6,704}  & 26.9 & 4.9  \\
& & (32) & (64) \\
\hline
\multirow{2}{*}{1208 Atoms}& \multirow{2}{*}{10,440}  & 47.3 & 7.1 \\
& & (48) & (96) \\
\hline
\end{tabular}}
\end{minipage}
\begin{minipage}{0.48\linewidth}
\centering
\subcaption{\textbf{Case Study:} Pt nanocluster}
\footnotesize{\begin{tabular}{|c|c|c|c|}
\hline
System  & Electrons & $\tau$-\qe (s)   & $\tau$-\pawfe (s)   \\
 -& - & (\# nodes) & (\# nodes) \\
\hline
\multirow{2}{*}{Pt\textsubscript{309}} & \multirow{2}{*}{3,090}  & 6.0 &1.8 \\
& & (16) & (32) \\
\hline
\multirow{2}{*}{Pt\textsubscript{561}}  & \multirow{2}{*}{5,610} & 13.7 & 2.7 \\
& & (32) & (64) \\
\hline
\multirow{2}{*}{Pt\textsubscript{923}}  & \multirow{2}{*}{9,230} & 36.3 & 5.1 \\
& & (48) & (96) \\
\hline
\end{tabular}}
\end{minipage}
\caption{Minimum wall time per SCF iteration comparison between \pawfe~and \qe. The calculations are run on {ALCF Polaris} supercomputer (NVIDIA A100 GPUs). The benchmark systems considered are (a) various sizes of WS\textsubscript{2} slab with Te atoms deposition (b) Pt nanocluster of various sizes. The table shows the number of electrons, the minimum wall time in seconds in both methods and the number of nodes (\# nodes) that was used to run the calculation.}
\label{tab: Min Wall Time}
\end{table}

\subsubsection{Comparison against DFT-FE with ONCV Pseudopotentials}
\begin{figure}[H]
    \centering
    \begin{subfigure}{0.48\textwidth}
        \centering
        \includegraphics[width=1.0\linewidth]{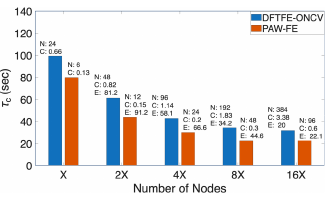}
        \caption{\centering \textbf{Case study}: OLCF Frontier}
    \end{subfigure}%
    \begin{subfigure}{0.48\textwidth}
        \centering
        \includegraphics[width=1.0\linewidth]{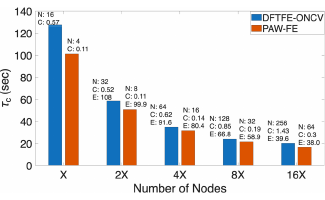}
        \caption{\centering \textbf{Case study}: ALCF Aurora}
    \end{subfigure}
    \caption{\small{Performance comparison of \DFTFE~ONCV pseudopotential calculation against \pawfe.  The super-computing systems considered are: (a) OLCF Frontier and (b) ALCF Aurora. The plots shows the average time per SCF iteration ($\tau_c$) in seconds on the vertical axis and the number of nodes for each calculation on horizontal axis where X denotes the minimum nodes required to run the calculation. The text inset describes the number of nodes (N), computational cost ($\eta_c$) in Node-hrs ($\eta_c$) and parallel scaling efficiency in \% (E).}}
    \label{fig:ComparisonONCVPAW}
\end{figure}

\begin{figure}[H]
    \centering
    \begin{subfigure}{0.48\textwidth}
        \centering
        \includegraphics[width=1.0\linewidth]{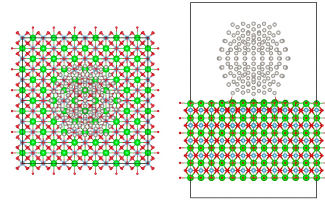}
        \caption{\centering \textbf{ONCV-PAW comparison system}: Pt\textsubscript{309} deposited on 4 layers of SrTiO\textsubscript{3}}
    \end{subfigure}%
    \begin{subfigure}{0.48\textwidth}
        \centering
        \includegraphics[width=1.0\linewidth]{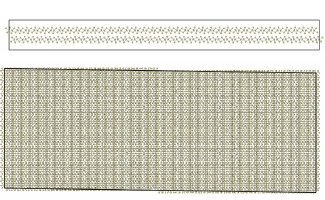}
        
        \caption{\centering \textbf{Large-scale calculation system}: Twisted bilayer WTe\textsubscript{2}, twist angle of $1.69^\circ$}
    \end{subfigure}
    \caption{ \textbf{Performance benchmarking systems:} Atomic structure of the systems considered for (a) Comparison against DFT-FE with ONCV pseudopotential (b) Leveraging exascale resources for density functional theory calculation. }
    \label{fig:Structures for performance BM}
\end{figure}

In this subsection, we compare the computational cost and parallel scaling efficiency of the proposed computational methodology of \pawfe~with the norm-conserving pseudopotential (ONCV) calculations~\cite{ONCV} using the \DFTFE~code\cite{dftfe0.6,dftfe1.0} solved to a similar level of accuracy. \cb \DFTFE, which has demonstrated exceptional parallel scalability on large-scale GPUs~\cite{Das2019FastSystem, Das2023}, provide a stringent benchmark against which the scalability of \pawfe~can be assessed. A key objective of the methods developed in this work is to ensure that the additional computational and communication overheads arising from the PAW formalism do not adversely affect the parallel scaling behaviour. To this end, in this benchmarking study, both \DFTFE~and \pawfe~leverage mixed-precision computation and communication, along with compute-communication overlap, to achieve high performance on modern heterogeneous architectures\cn. The ONCV pseudopotential is obtained from SG15\cite{ONCV} repository. As alluded to earlier, these DFT calculations employing ONCV pseudopotentials can be very challenging, requiring a large number of DoFs to achieve the desired accuracy for most material systems. 
The strong scaling behavior of \DFTFE~using ONCV pseudopotentials and \pawfe~is assessed in terms of computational cost, parallel efficiency, and wall time per SCF iteration. Figure~\ref{fig:ComparisonONCVPAW} presents the average wall time per SCF iteration ($\tau_c$) as a function of the number of compute nodes on ALCF Aurora and OLCF Frontier. In this study, we consider the system of Pt\textsubscript{309} deposited on SrTiO\textsubscript{3} slab. The system comprises 2253 atoms with 21,114 electrons using ONCV pseudopotential and 18,642 electrons using PAW dataset. 
Figure~\ref{fig:Structures for performance BM}(a) shows the structure considered for the comparison. 
The FE-discretization parameters that reach the desired discretization error in energy and forces are mentioned in column four of Table~\ref{tab:fe_parameters_pawfe} for \pawfe~resulting in about $5,000~\frac{\text{DoFs}}{atom}$, while for the ONCV calculation we use $p = 7, p_{\text{el}} = 7, h = 1.17$ with close to $20,000~\frac{\text{DoFs}}{atom}$. 
From the plots, we observe that \pawfe~requires approximately 4$\times$ fewer compute nodes than \DFTFE, with 5$\times$--6$\times$ lower computational cost, consistent with the expected $\mathcal{O}(MN_e)$ memory scaling and the ratios of basis functions and number of electrons. Furthermore, we observe that \pawfe~is $5\times$ more efficient than \DFTFE~with ONCV pseudopotentials on both ALCF Aurora and OLCF Frontier. The efficiency gains obtained in \pawfe~can be attributed to (i) coarser finite-element mesh size leading to far fewer degrees of freedom compared to the ONCV pseudopotentials used in \DFTFE~calculations to reach the desired \cb energy and force discretization error \cn, (ii) reduced Chebyshev polynomial degree during the ChFSI step due to the coarser meshes in \pawfe~calculations, and  (iii) fewer valence electrons within the frozen-core approximation of the PAW method in the case of Pt, leading to better parallel scaling efficiency of ChFSI due to reduced communication costs in the Rayleigh-Ritz(RR) step. We note that the computational gains of \pawfe~relative to \DFTFE~using ONCV pseudopotentials may depend on the choice of pseudopotential repository or PAW dataset. Most existing PAW datasets are developed for plane-wave implementations and typically restrict the number of partial waves (${\phiae}, {\phips}$) to two per angular momentum quantum number. Increasing the number of partial waves can reduce the degrees of freedom required to achieve desired accuracies without incurring additional computational cost in \pawfe, and this is part of our future work. Additionally, we observe that the scalability of \pawfe~is comparable to that of \DFTFE, which is well known for its strong parallel scalability~\cite{Das2019FastSystem,Das2023}. This highlights the effectiveness of the computational methodologies developed for solving the generalized eigenproblem arising in \pawfe.

The preceding results demonstrate that the proposed GPU-centric algorithms substantially reduce both computational cost and resource requirements. We now showcase these capabilities through a large-scale \textit{ab initio} calculation that would otherwise be prohibitively expensive using conventional approaches.
\subsubsection{Large-scale Density Functional Theory Calculations}
\cb To demonstrate the large-scale capability of \pawfe, we consider a twisted bilayer WTe\textsubscript{2} system as a representative benchmark. Twisted bilayer materials are computationally demanding because the large moiré supercells generated by small twist angles can contain tens to hundreds of thousands of electrons, making them an ideal testbed for large-scale first-principles calculations. The benchmark system comprises two WTe\textsubscript{2} layers separated by an interlayer distance of 4.21 bohr and twisted by an angle of $1.69^{\circ}$, as shown in Fig.~\ref{fig:Structures for performance BM}(b). The resulting system comprises 11,544 atoms and 130,832 electrons. The finite-element discretization parameters used in this study are reported in the last column of Table~\ref{tab:fe_parameters_pawfe}, corresponding to approximately 6300 DoFs per atom.
\cn

As summarized in Table~\ref{tab:largeSystemTime}, the calculation requires a minimum of 100 nodes (1,200 Intel GPUs) on ALCF Aurora, with a wavefunction block size of 150 employed in both the Chebyshev filtering and Rayleigh-Ritz steps. At this minimum configuration, the computational cost is approximately 10 node-hours (7 minutes) per SCF iteration. Increasing the allocation to 200 nodes allows us to use a wavefunction block size of 800 and reduces the SCF iteration time to approximately 3 minutes, while using 400 nodes further reduces it to about 2 minutes. Beyond this regime, the parallel efficiency is limited by the current scalability of the \texttt{oneCCL} library used for collective communication.

\begin{table}
    \centering
    \begin{tabular}{|c|c|c|c|c|}
      \hline
       \# Nodes  & $\frac{\text{DoFs}}{\text{GPU}}$  & SCF Time (s) & CF time (s)  & RR step (s)   \\
       \hline
     100    &  60,586 &  429.50 & 89.36 &  311.30 \\
     \hline
      200   &  30,293  & 178.78 & 35.79 & 125.60  \\
      \hline
      400   &  15,147  & 118.75 & 21.83 &  82.61 \\
      \hline
    \end{tabular}
    \caption{Bilayer WTe\textsubscript{2} SCF time, Chebyshev filtering (CF) and Rayleigh Ritz (RR) step time in seconds on ALCF Aurora supercomputer. }
    \label{tab:largeSystemTime}
\end{table}

%% file: Discussion.tex
In this work, we have presented the key computational methodologies underlying \pawfe, a finite-element discretized projector augmented-wave DFT formulation designed for accurate, efficient, and scalable electronic-structure calculations on modern multi-node GPU architectures. The proposed approach 
is built around a hardware-aware eigensolver strategy that leverages the approximate inverse of the PAW overlap matrix, mixed-precision arithmetic, low-precision communication, and computation–communication overlap. Relative to established plane-wave PAW based DFT codes, \pawfe~substantially reduces time-to-solution across a wide range of system sizes while naturally accommodating generic boundary conditions, and demonstrates scalability to the $10^4$--$10^5$ electron regime otherwise inaccessible to existing systematically convergent DFT implementations.

A central design principle of~\pawfe~is the synergy between the PAW formalism and high-order finite-element discretization. In particular, the multi-resolution quadrature rules employed in the finite-element framework enable accurate evaluation of atom-centered PAW integrals without increasing the number of degrees of freedom, preserving the efficiency of the discretized representation. Further, the locality of the FE basis induces a sparse operator structure requiring only nearest-neighbor communication, a property that maps favorably onto the distributed memory hierarchies of exascale architectures and naturally accommodates periodic, semi-periodic, and non-periodic boundary conditions within a single unified formulation. The latter capability enables direct, artifact-free modeling of slabs, surfaces, nanoclusters, and interfaces that would otherwise require artificial periodicity or large vacuum padding in widely used plane-wave approaches.

To achieve high computational efficiency on modern architectures, the operations within each SCF iteration are reformulated as high-arithmetic-intensity matrix–matrix multiplications well-suited to GPU execution. The dominant computational step, the solution of the PAW generalized Hermitian eigenproblem (PAW-GHEP), is addressed through the Residual-based Chebyshev Filtered Subspace Iteration (R-ChFSI) algorithm. R-ChFSI, being tolerant to approximations in matrix–multivector products arising in the construction of the desired eigensubspace, admits several performance-oriented strategies without compromising the accuracy of the converged solution. Specifically, a computationally inexpensive FE-discretized PAW overlap matrix inverse is introduced; reduced-precision arithmetic (FP32/TF32) is employed during Chebyshev-filtered subspace construction; nearest-neighbor data transfers are performed in BF16 to reduce communication volume by a factor of 4$\times$ relative to FP64 during filtering; and a block-wise overlap strategy is developed for orchestrating computation and communication throughout.  Together, these strategies yield a 3$\times$ reduction in wall time for Chebyshev-filtered subspace construction and an overall CPU--GPU speedup of 8$\times$--20$\times$ across Intel and AMD GPU architectures. \pawfe~achieves a 7$\times$ reduction in minimum wall time to solution relative to plane-wave PAW codes for systems of approximately 10,000 electrons, with greater gains expected for larger system sizes and a speedup of up to 5$\times$ relative to the ONCV-based \DFTFE~implementation. Time-to-solution per SCF iteration of just 2 minutes for systems comprising 130,000 electrons further demonstrates the practical scalability of \pawfe~to realistic large-scale simulations.

Looking ahead, the computational framework developed here opens several avenues for further development. A natural next step is the incorporation of kinetic-energy-dependent meta-GGA functionals such as \texttt{r$^2$SCAN}~\cite{r2scan} and \texttt{LAK}~\cite{LAK}, and hybrid functionals, which occupy higher rungs of Jacob's ladder of exchange-correlation approximations. Incorporating solvent effects within the present framework is also of immediate interest for catalytic and electrochemical applications involving nanometer-scale systems. Extending the current collinear spin formalism to non-collinear magnetism and relativistic effects will be important for accurate modeling of magnetic materials and heavy elements and will form part of future investigations. Finally, systematic guidelines for the selection of finite-element discretization parameters $p$, $p_{\mathrm{el}}$, and $h$, together with appropriate quadrature rules, particularly for the \texttt{JTHv2.0}~\cite{JTH_PBE_v2_0} PAW datasets, will be provided in forthcoming work to facilitate broader adoption of \pawfe~across diverse materials systems.

%% file: acknowledgement.tex
 This research used computational resources at the Argonne and the Oak Ridge Leadership Computing Facilities. The Argonne Leadership Computing Facility at Argonne National Laboratory (ANL) is supported by the Office of Science of the U.S. DOE under Contract No. DE-AC02-06CH11357. The Oak Ridge Leadership Computing Facility at the Oak Ridge National Laboratory is supported by the Office of Science of the U.S. DOE under Contract No. DE-AC05-00OR22725. The research also used the PARAM Pravega resources at the Indian Institute of Science, supported by the National Supercomputing Mission (NSM). P.M acknowledges Vishwas Rao from ANL for useful discussions on the manuscript. K.R acknowledges Nishant Gupta for support on reduced precision communication strategies employed in the current work. P.M. acknowledges financial support from the Google India Research Award 2023 and ARG MATRICS (Grant No: ANRF/ARGM/2025/002824/TS) from Anusandhan National Research Foundation (ANRF) during the course
of this work. P.M also acknowledges funding support from the National Quantum Mission, an initiative of the Department of Science and Technology, India. P.M also acknowledge the support provided by the foundation of QC Innovation (FQCI), DST-NQM T-hub at IISc, in facilitating this project. K.R. acknowledges the Prime Minister Research Fellowship (PMRF) from the Ministry of Education India for financial support. 